\documentclass[preprint,aps,prb,showpacs,superscriptaddress,floatfix,nofootinbib,ruledtabular]{revtex4-1}

\usepackage{graphicx}
\usepackage{bbold} 
\usepackage{dsfont}
\usepackage{multirow}
\usepackage[section]{placeins}
\usepackage{color}
\usepackage{ulem}
\usepackage[caption=false]{subfig}
\usepackage{float}
\usepackage{color} 
\usepackage{verbatim}
\usepackage{subfig}
  
\usepackage{amssymb} 
\usepackage{amsmath} 
\usepackage{wasysym}
\usepackage[utf8]{inputenc} 

\begin{document}


\title{Topological Triplon Modes and Bound States in a Shastry-Sutherland Magnet}


\author{P. A. McClarty} 
\affiliation{ISIS, Science and Technology Facilities Council, Rutherford Appleton Laboratory, Didcot OX11 0QX, United Kingdom.}

\author{F. Kr\"uger} 
\affiliation{ISIS, Science and Technology Facilities Council, Rutherford Appleton Laboratory, Didcot OX11 0QX, United Kingdom.}
\affiliation{London Centre for Nanotechnology, University College London, Gordon Street, London, WC1H 0AH, United Kingdom.}

\author{T. Guidi} 
\affiliation{ISIS, Science and Technology Facilities Council, Rutherford Appleton Laboratory, Didcot OX11 0QX, United Kingdom.}

\author{S.~F. Parker} 
\affiliation{ISIS, Science and Technology Facilities Council, Rutherford Appleton Laboratory, Didcot OX11 0QX, United Kingdom.}

\author{K. Refson} 
\affiliation{ISIS, Science and Technology Facilities Council, Rutherford Appleton Laboratory, Didcot OX11 0QX, United Kingdom.}
\affiliation{Department of Physics, Royal Holloway, University of London, Egham TW20 0EX, United Kingdom.}

\author{A.~W. Parker} 
\affiliation{Central Laser Facility, Science and Technology Facilities Council, Rutherford Appleton Laboratory, Didcot OX11 0QX, United Kingdom.}

\author{D. Prabakharan} 
\affiliation{Clarendon Laboratory, University of Oxford, Parks Road, Oxford OX1 3PU, United Kingdom.}

\author{R. Coldea} 
\affiliation{Clarendon Laboratory, University of Oxford, Parks Road, Oxford OX1 3PU, United Kingdom.}

\newcommand{\scbo}{SrCu$_{2}$(BO$_{3}$)$_2$}


\date{\today}

\maketitle


{\bf The twin discoveries of the quantum Hall effect,\cite{Klitzing1980}  in the 1980's, and of topological band insulators,\cite{Hsieh2008}  in the 2000's, were landmarks in physics that enriched our view of the electronic properties of solids. In a nutshell, these discoveries have taught us that quantum mechanical wavefunctions in crystalline solids may carry nontrivial topological invariants which have ramifications for the observable physics. One of the side effects of the recent topological insulator revolution has been that such physics is much more widespread than was appreciated ten years ago.  For example, while topological insulators were originally studied in the context of electron wavefunctions,  recent work has led to proposals of topological insulators in bosonic systems: in photonic crystals,\cite{Rechtsman2013} in the vibrational  modes of crystals,\cite{Zhang2010} and in the excitations of ordered magnets.\cite{Katsura2010}  Here we confirm the recent proposal\cite{Romhanyi2015} that, in a weak magnetic field,  the dimerized quantum magnet SrCu$_{2}$(BO$_{3}$)$_2$ is a bosonic topological insulator with nonzero Chern number in the triplon bands and topologically protected chiral edge excitations. }

In this letter,  we carry out a detailed examination of the original theoretical proposal.\cite{Romhanyi2015}  We present new inelastic neutron scattering results exploring the triplon bands in small magnetic fields of up to $2.8$ T perpendicular to the dimer planes which provides unprecedented insight into the nature of the magnetic couplings in this material. In addition to the single triplon modes, we find a new comparatively dispersive feature that hybridizes with them. We present a detailed theoretical scenario that accounts for the presence of such a feature in the data in terms of a singlet bound state of two triplons.\cite{Totsuka2001} On the basis of this revised model for the low energy excitations, we make predictions for the presence of multiple topological transitions, the thermal Hall effect, and the presence of edge states.

The quantum magnet \scbo\  is famous in the magnetism community\cite{Gozar2005} especially for its rich in-field phase diagram reflected in a series of magnetization plateaus.\cite{Kodama2002,Corboz2014}  The material is composed of layers of strongly interacting  $S=1/2$  copper moments that bind together in pairs (dimers), forming quantum mechanical singlets. Neighboring dimers have an orthogonal arrangement (Fig.~1).  Most magnetic materials undergo a transition into long-range magnetic order so the fact that the ground state of this material is both interacting and with only short-range correlations is remarkable: a consequence of the frustrating effect of the Shastry-Sutherland lattice geometry.\cite{Shastry1981,Miyahara1999}  The lattice geometry of \scbo\ is also responsible for ensuring that the excited states of the magnet - called triplons - are almost flat across the Brillouin zone.\cite{Kageyama2000,Gaulin2004,Aso2005}  The predominant contribution to the weak dispersion of these modes is due to subleading magnetic exchange couplings which are antisymmetric Dzyaloshinskii-Moriya (DM) interactions. \cite{Cepas2001,Romhanyi2011}  These DM interactions are responsible for complex hopping amplitudes of the triplons which may then pick up Berry phases around closed paths. Their role is therefore similar to that of spin-orbit coupling in electronic topological insulators.  The heart of the aforementioned proposal \cite{Romhanyi2015} is that the application of a small magnetic field leads to triplon bands acquiring a nontrivial topological invariant called a Chern number which implies the existence of chiral magnetic edge states. 

\begin{figure}[htpb]
\centering
\hspace{3cm}
\centering
\includegraphics[width=0.85\columnwidth]{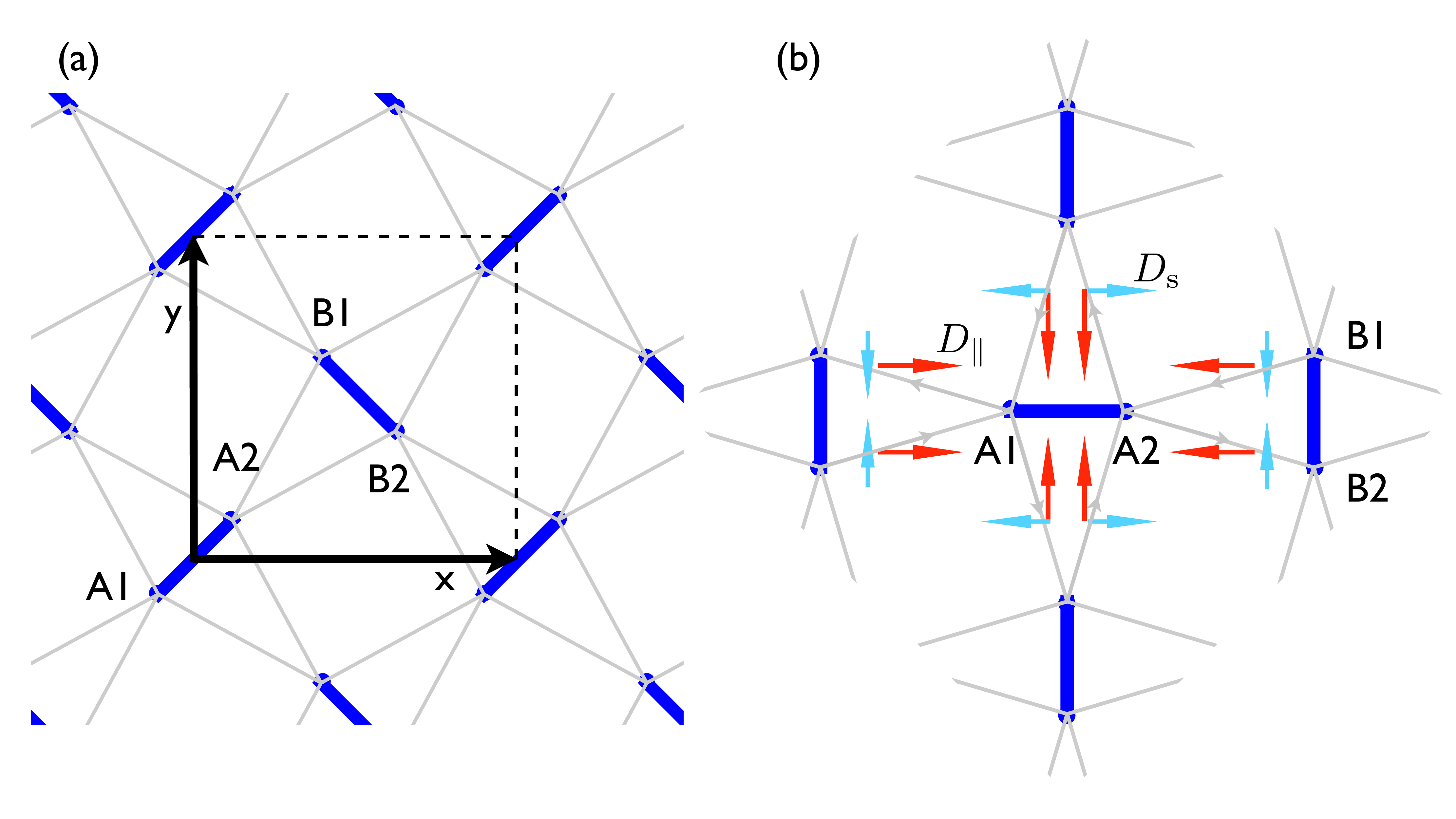}
\caption{ 
{\bf Symmetries and exchange on the magnetic lattice of \scbo.} (a) The copper-copper singlets are indicated by the thick blue lines which are coupled by exchange coupling $J$. Neighboring singlets joined by light grey bonds are coupled by $J'$ exchange. The crystallographic unit cell is the dashed square. (b) Conventions and symmetries of the inter-dimer DM exchange. The figure shows the staggered and 
parallel in-plane components, $D_s'$ and $D_\parallel'$. In addition, there exists a uniform component $D_\perp'$ perpendicular to the plane.
\label{fig:S1}}
\end{figure}


Crystals were grown with 99\% enriched boron-11 by the optical floating zone technique.\cite{Smith1991} A large crystal of size $4$ cm was grown at $0.25$ mm/h under $3$ bar oxygen pressure. The neutron experiment was carried out using three pieces of this crystal of total mass $5.9$ g which were co-aligned on the ALF instrument at ISIS on an aluminum mount. 

Our inelastic neutron scattering (INS) measurements were performed using the direct geometry time-of-flight spectrometer, LET, at the ISIS facility.\cite{LET} The sample was mounted inside a $9$ T superconducting magnet and the sample was cooled down to $2$ K. The measurements were performed with multiple incident energies of which we focus here on the $5$ meV  data. We collected multi-angle Horace \cite{Horace} scans with $1^{\circ}$ step sizes at $0$ T, $0.7$ T and $1.4$ T and $2^{\circ}$ step sizes at $2.8$ T with a $78^{\circ}$ total coverage. The experimental resolution at $E_i =5$ meV was measured at the elastic line (FWHM) and the calculated energy resolution at $3$ meV was then calculated to be $0.059$ meV.


\begin{figure}[htpb]
\centering
\hspace{3cm}
\centering
\includegraphics[width=\columnwidth]{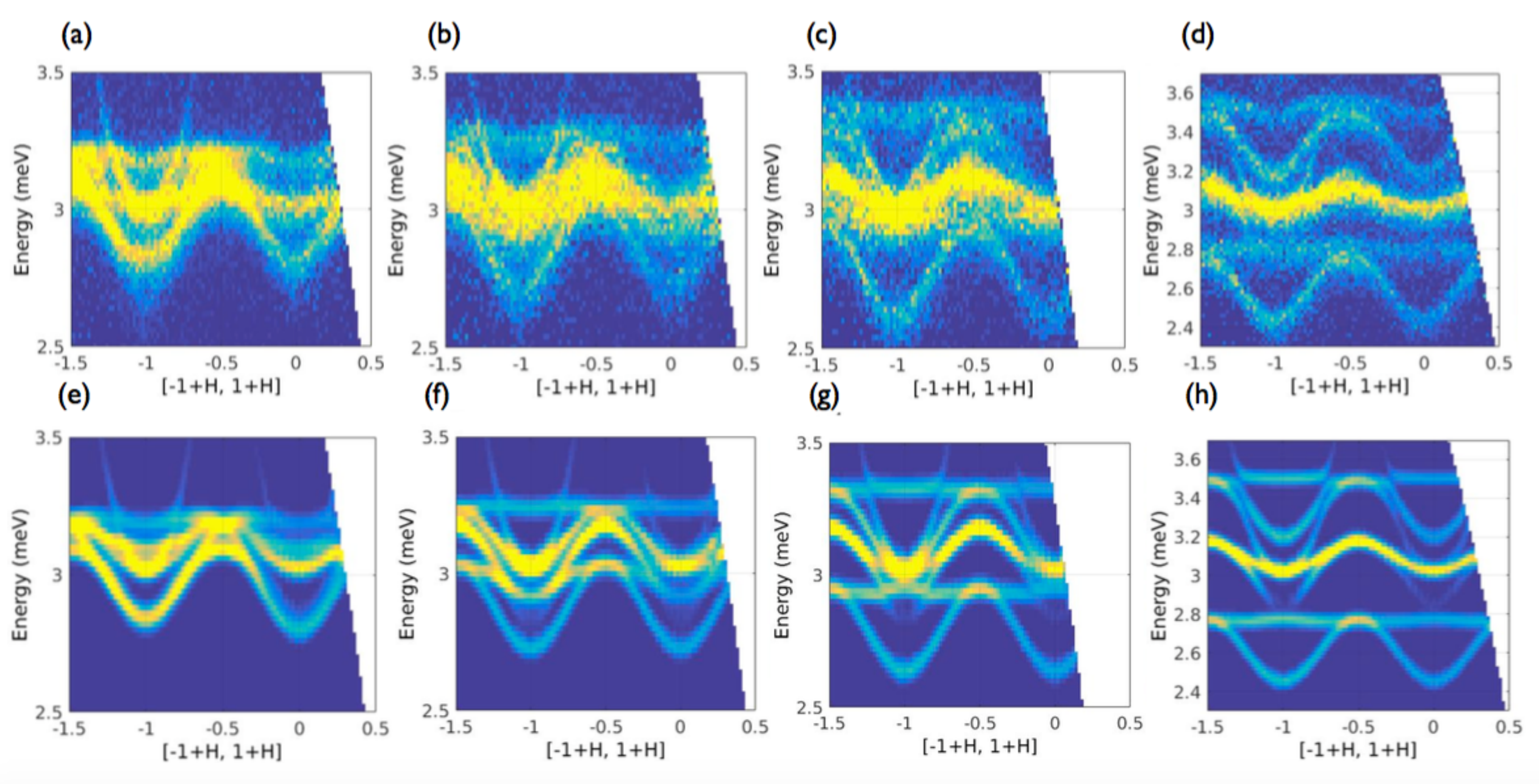}
\caption{ 
{\bf Evolution of the single triplon modes in a field along the $[001]$ direction with cut taken in the $[-1+H,1+H]$ direction.} Panels (a) and (e) respectively show zero field data and the calculated dynamical response, with (b) and (f) at $0.7$ T, (c) and (g) at $1.4$ T and (d) and (h) at $2.8$ T. These cuts were integrated in the $L$ direction between $\pm 0.2$ and in other directions between $\pm 0.1$.
\label{fig:M1}}
\end{figure}

Fig.~\ref{fig:M1}(a)-(d) shows INS cuts along the $[-1+H,1+H]$ direction showing the single triplon excitations. At zero field, the gap to the triplons is roughly $3$ meV which is the scale of the nearest neighbor isotropic exchange $J$. The non-degeneracy of these modes implies the presence of anisotropic exchange. In order to understand the 
experimental dispersion, we consider all symmetry-allowed couplings between first and second neighbor spins. There are four allowed intra-dimer exchange couplings -- the isotropic exchange $J$, a Dzyaloshinskii-Moriya (DM) coupling $D$ with $\boldsymbol{D}$ vector in the $c$ direction, an Ising coupling $J_{\rm zz}$, and a symmetric exchange $J_{\rm xy}$ -- and the corresponding 
single-dimer Hamiltonian is given by
\begin{equation}
\mathcal{H}_0 = J \boldsymbol{S}_1\boldsymbol{S}_2+\boldsymbol{D}\cdot\left( \boldsymbol{S}_1\times \boldsymbol{S}_2  \right)+J_{\rm zz} S^{\rm z}_1 S^{\rm z}_2\pm
 J_{\rm xy}\left(S^{\rm x}_1 S^{\rm y}_2 + S^{\rm y}_1 S^{\rm x}_2\right)-g_{\rm z}\mu_{\rm B}B\left(  S^{\rm z}_1 + S^{\rm z}_2 \right),
\end{equation}
where we have included the Zeeman term for a field perpendicular to the Shastry-Sutherland planes. Note that by symmetry, the $J_{\rm xy}$ coupling is of opposite sign on the two sub-lattices.  Owing to the weak spin-orbit, we {\it a priori} expect the hierarchy of energy scales $J > D > J_{\rm zz},J_{\rm xy}$.  We use the bond operator formalism\cite{Sachdev1990,Cheng2007} to represent the dimer spin Hamiltonian in terms of singlet and triplet operators, $S_{1/2}^{\alpha}  = \frac{1}{2} ( \pm s^{\dagger} t_{\alpha} \pm  t_{\alpha}^{\dagger}s - i\epsilon_{\alpha\beta\gamma} t_{\beta}^{\dagger} t_{\gamma} )$. The operators $s^\dagger$ and $t_\alpha^\dagger$ ($\alpha=x,y,z$) create 
dimer singlet and triplet states out of the vacuum, respectively, and are subject to the usual hard-core constraint $s^\dagger s+\sum_\alpha t_\alpha^\dagger t_\alpha=1$. The coupling between dimers gives dynamics to the triplon excitations.  Again, the largest such coupling between next-nearest neighbor spins is the isotropic component $J'$ of the Heisenberg exchange, followed by the DM coupling $\boldsymbol{D}'$. There are no symmetry constraints on the components of $\boldsymbol{D}'$ on a bond but, once those are fixed, they are determined over the entire lattice [see Fig.~1(b)].

We condense the singlets into the ground state and perform a unitary rotation to eliminate linear terms in the triplet operators arising from the intra-dimer DM interaction D.  The 
triplon dispersions are computed by diagonalizing the quadratic triplon Hamiltonian (Supplementary Information). In the Hamiltonian we only keep DM terms to linear order. 
It turns out that only two of the three components of the inter-dimer DM exchange, $D_\perp'$ and $D_\parallel'$, contribute.   It is well known \cite{Romhanyi2015} that a dispersion in the single triplons coming from $J'$ (in the absence of DM interactions) arises only to sixth order in $J'/J$. We neglect this lowest order contribution to the triplon hopping. The anomalous terms that arise in the bond-operator expansion only give a negligible correction to the 
triplon dispersion (Supplementary Information).  In summary, we obtain six triplon bands since there exist two dimers in the unit cell. These triplon excitations depend on five parameters ($J$, $J_{\rm zz}$, $J_{\rm xy}$, $D_\perp'$, and $D_\parallel'$). Following Ref.~[\onlinecite{Romhanyi2015}], we also include a small triplon hopping term between dimers on the same sublattices.


We take constant $\boldsymbol{k}$ cuts through the data and fit the peaks to gaussians with variable mean and variance taking the minimal number of gaussians necessary to obtain a good fit to each cut. In this way we obtain a set of points tracking the dispersion curves of the single triplon modes (Supplementary Information). To these points, we fit the single triplon model using a least square minimization algorithm and obtain a set of exchange parameters. 

\begin{figure}[htpb]
\centering
\includegraphics[width=0.65\columnwidth]{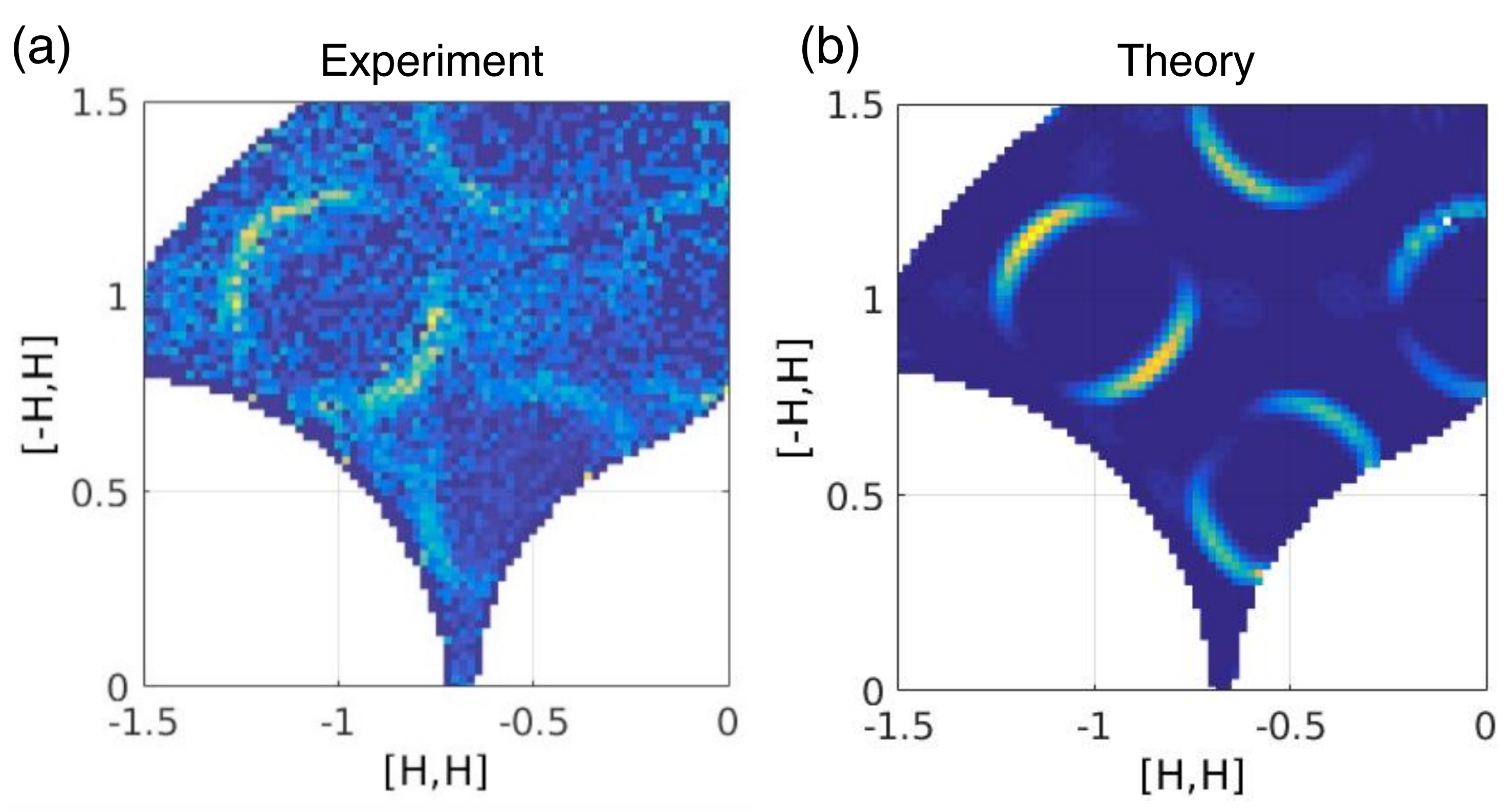}  
\caption{{\bf Mode X at constant energy} (a) A constant energy cut showing rings of intensity in the $[HH]/[-HH]$ plane integrated over the range $3.25-3.35$ meV and over the measured $[00L]$ direction at zero field. (b) The calculated dynamical structure factor for comparison with the experiment.
\label{fig:M2}}
\end{figure}

In addition to the single triplon excitations,  Fig.~\ref{fig:M1} shows a more dispersive mode that intersects them with a minimum at the $\Gamma$ point  at around 
$3$ meV, which is roughly in the middle of the triplon bands. This feature was not apparent in earlier inelastic neutron-scattering data (taken in zero field).\cite{Gaulin2004}
The intensity of this mode (hereafter called mode X) just above the triplon modes is around three percent of the maximum intensity of the single triplons and can be observed using LET because of the high sensitivity of the spectrometer. 
The additional mode hybridizes with the single triplon modes - the hybridization gap being most visible at higher fields. A constant energy cut is shown in Fig.~\ref{fig:M2} at $3.3$ meV showing rings 
of intensity coming from mode X which meet the Brillouin zone edge at about $3.4$ meV. An applied field has no apparent effect on the dispersion relation of X (Supplementary Information).   Two possibilities for the identity of mode X present themselves: one is that it is a bound state of triplet modes with net angular momentum zero and the second is that it is a phonon.    

The bond operator representation leads to cubic and quartic interaction terms between the triplons which are responsible for a tower of bound states \cite{Gozar2005}. In order to study the bound states in the singlet sector as motivated by the experiment, we follow and extend the results of Ref.~\onlinecite{Totsuka2001}. A singlet bound state of two triplons has a wavefunction $\vert \Phi_{\boldsymbol{K}}  \rangle = \frac{1}{\cal{N}} \sum_{\boldsymbol{q},\alpha} \Phi_{\boldsymbol{K},\boldsymbol{q}}  t^\dagger_{\frac{\boldsymbol{K}}{2}+\boldsymbol{q},\alpha} t^\dagger_{\frac{\boldsymbol{K}}{2}-\boldsymbol{q},\alpha}  \vert 0\rangle$.  We approximate this wavefunction by carrying out degenerate perturbation theory in $J'$ to third order within the two triplon sector (Supplementary Information).  This generates nearest and next-nearest neighbor potentials and effective hopping terms and requires us to consider eight localized two-triplon states. There is no linear contribution of the DM interactions to the bound-state sector. As for the single triplons we neglect higher order DM terms.
It turns out that, to third order in $J'$, two pairs of four such states decouple leading to degeneracies that are artifacts of the low order perturbation theory. Therefore, we include those terms to fourth order in $J'$ that are necessary to couple these sectors. The effective Hamiltonian for the singlet bound state depends, to this order, only on the bare $J$ and $J'$ exchange couplings. Diagonalization of this Hamiltonian leads to four bound states and four anti-bound states. The lowest energy bound state is at the Brillouin zone centre as observed experimentally (see Fig.~\ref{fig:M1}). 

The spin Hamiltonian described above -- which is the minimal model necessary to describe the single triplon and $S=0$ two triplon sectors -- provides a natural explanation for the existence of a 
significant  hybridization between these sectors. In the bond-operator language, the DM interaction gives rise to a cubic triplon
term that {\it linearly} couples the bound state and single-triplon sectors.  Whereas only two of the three inter-dimer DM couplings contribute to the single triplon hopping Hamiltonian, the hybridization Hamiltonian includes all three (Supplementary Information). Once the bound state mixes some single triplon character we also understand how the bound state acquires some neutron scattering intensity given that the singlet sector on its own and the ground state are not coupled by neutrons.\cite{remark} 
 
The above method may be used to study $S=1$ and $S=2$ two-triplon bound states. \cite{Totsuka2001} Within the model these naturally lie at higher energies than the lowest energy singlet sector. And, indeed, it appears that the lowest energy $S=1$ bound states in \scbo\ lie just below the two-triplon continuum. 


We have seen that the model we studied for the single triplons straightforwardly allows for the existence of singlet bound states and their coupling to single triplon states. We now consider the validity of this model as an explanation for our INS results on \scbo. We extracted the dispersion of mode X from the data and fit the decoupled bound state to it. Then we allowed the single triplon and hybridization parameters to relax to their optimal values on the basis of the triplon dispersions obtained at $0$ T and $2.8$ T. This way we completely parameterize our Hamiltonian using constraints from the data. The resulting fits are shown in Fig.~\ref{fig:M1} for four different field strengths and for a cut in the $[H,0]$ direction. In the supplementary section comparisons for four other cuts are shown. 


Our model provides an excellent description of the inelastic neutron scattering data and can now be used to investigate the topological nature of the magnetic excitations. Already 
on the level of the single triplon bands we find significant differences compared to the model used in the original theoretical proposal.\cite{Romhanyi2015}  (i) The central triplon bands of \scbo\ are much 
more dispersive and, in the regime of small fields, their bandwidth overlaps significantly with those of the upper and lower triplon bands. (ii) In zero field [Fig.~1(a)]  there exist no Dirac points in the spectrum
and instead we find small gaps, which are due to the presence of small anisotropic intra-dimer exchanges $J_{\rm zz}$ and $J_{\rm xy}$. In addition, we have clear evidence of a two-triplon singlet bound state that hybridizes with the triplon bands.

In order to analyze whether, despite these complications, the magnetic excitations exhibit a non-trivial topological character, we calculate the first Chern number $C_n$ for each band $n$. This integer topological invariant is defined as an integral of the Berry curvature of the band over the two-dimensional Brillouin zone and can be calculated efficiently using the link variable method of Ref.~\onlinecite{Fukui2005}, taking special care of band-touching points (Supplementary Information).
The Chern number is connected to observable quantities. In integer quantum Hall systems, for example, it is related to the quantized Hall conductance.\cite{Thouless1982}
A bosonic analogue of this result for the case when the band is thermally populated is the thermal Hall effect which, rather than depending on the Chern number and yielding a quantized response instead depends on the magnitude of the Berry curvature in thermally occupied parts of the band.\cite{Katsura2010,Matsumoto2011} A topological transition is visible as a kink in the Hall effect resulting from a spike in the Berry curvature at the touching point between topologically non-trivial bands with equal and opposite Chern numbers. 

\begin{figure}[htpb]
\centering
\hspace{3cm}
\centering
\includegraphics[width=\columnwidth]{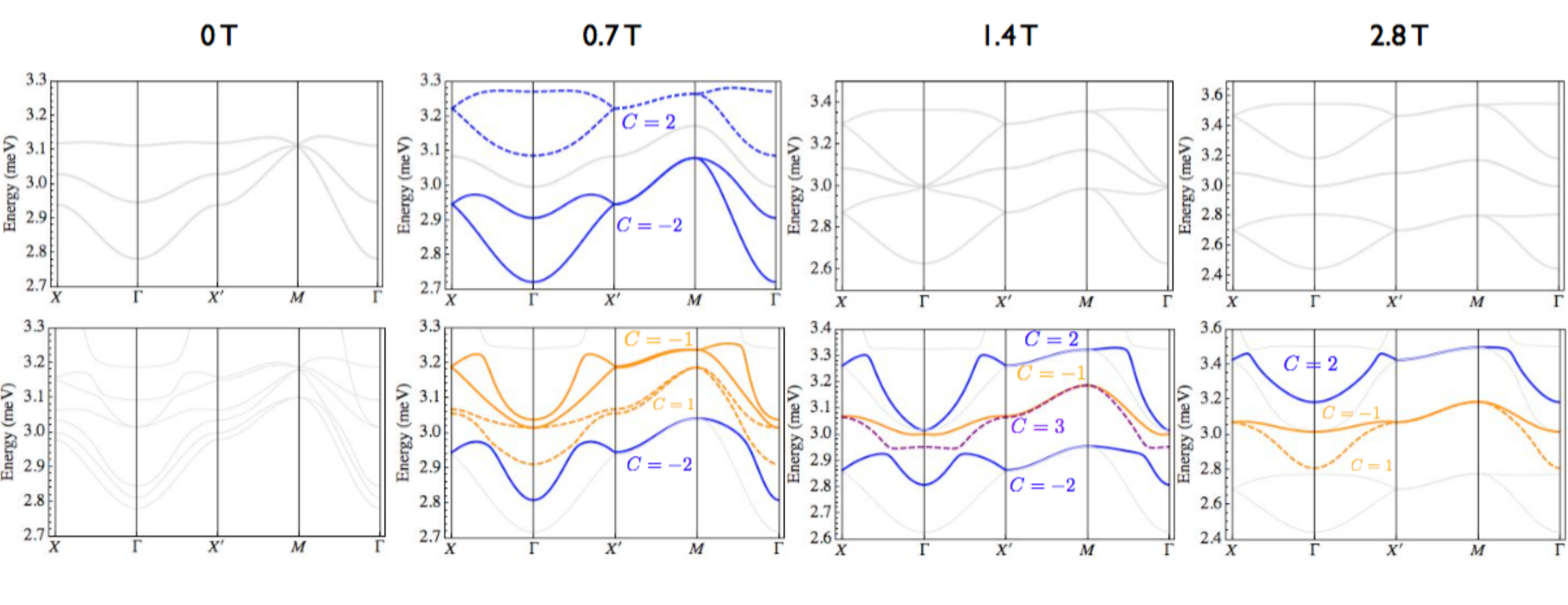}
\caption{
{\bf Chern numbers of triplon bands and lowest singlet bound state.} The upper figures show the prediction of the model of Ref.~\onlinecite{Romhanyi2015} but using the best fit parameters determined from our experiment.  The dispersions are shown along a high symmetry path.  Between $0$ T and $1.4$ T two of the singlet triplon bands have nontrivial Chern numbers $\pm 2$. A touching point at $1.4$ T leads to all bands being topologically trivial at higher fields. The lower figures show the predictions at four fields of the full model including the bound state. In contrast to the earlier work, we expect that Chern numbers of $\pm 1$ and $+3$ arise in the band structure. In addition, nontrivial bands persist to much higher fields $B>1.4$ T. Note that there are bands at higher energy (not shown) that ensure the total Berry flux across all modes is zero. 
\label{fig:Chern}}
\end{figure}

For the idealized triplon model of Romhanyi {\it et al.},\cite{Romhanyi2015} there exist only two topologicallly non-trivial triplon bands with Chern numbers $C=+2$ and $C=-2$ and a single
finite field topological transition at the upper critical field $B_c=1.4$ T. At $B_c$ these bands touch at the $\Gamma$ point and, as expected, the thermal Hall effects shows a kink. For our refined model the situation turns out to be much richer. Even without the bound state, the anisotropic exchanges $J_{\rm zz}$ and $J_{\rm xy}$ lead to the presence of two topological transitions. The hybridization with the bound state gives rise to a large number of bands with non-zero Chern numbers. Not surprisingly,  we therefore find a sequence of topological transitions involving different pairs of bands. This is also reflected by the extremely rich structure of the thermal Hall effects (Supplementary Information). Fig.~\ref{fig:Chern} shows the Chern numbers and dispersions of the low energy bands at four different fields 
for the idealized model of Romhanyi {\it et al.},\cite{Romhanyi2015} and our refined model that includes the hybridization with the bound state and provides an excellent description of the
measured magnetic excitations of \scbo. 

\begin{figure}[h!]
\includegraphics[width=0.8\columnwidth]{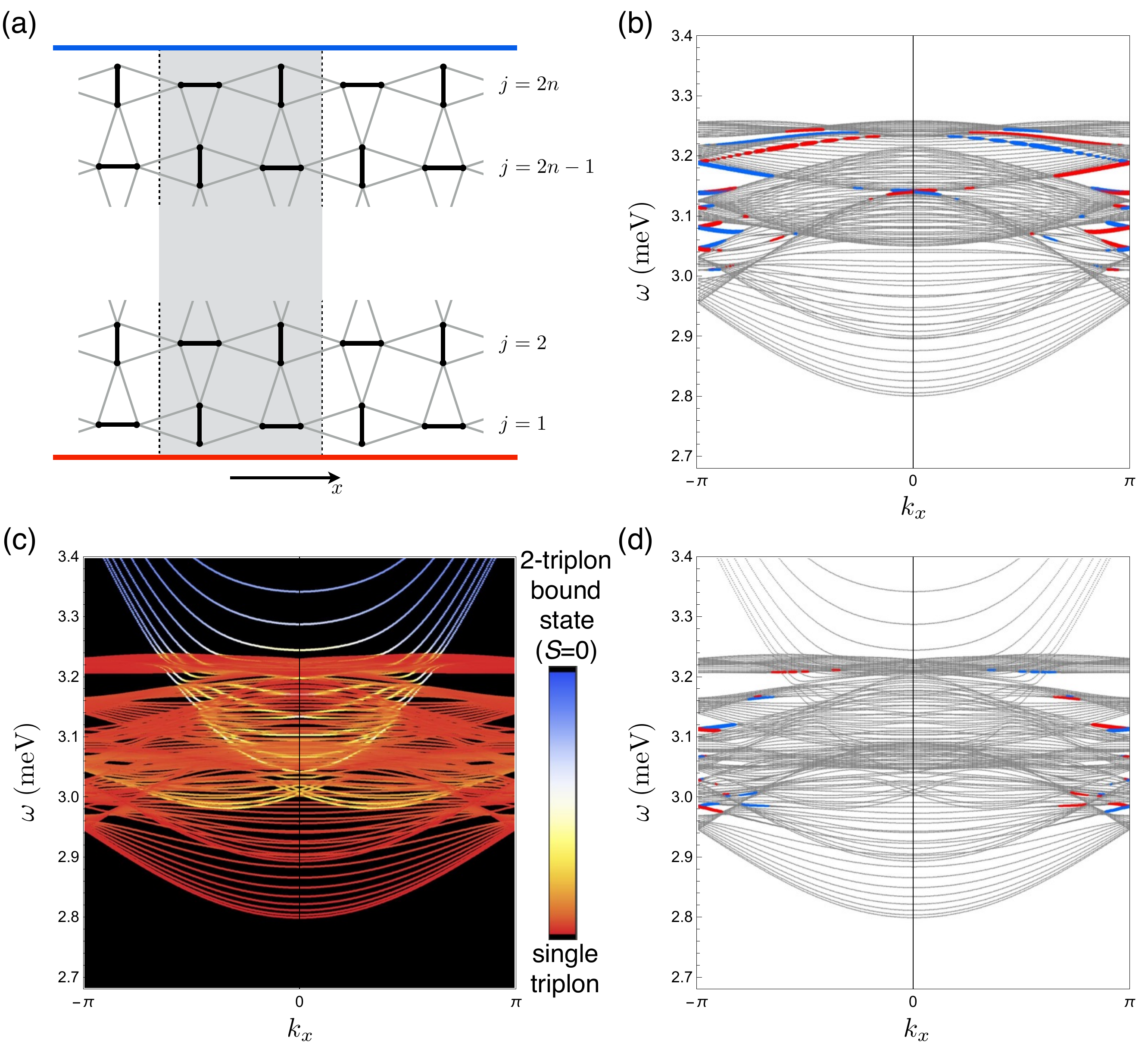}
\caption{
{\bf Edge states.} (a) Strip geometry used for our edge state calculation. (b) Excitation spectrum of a strip with 20 rows of dimers for the single triplon model (without bound state) in a field $B=0.35$ T
perpendicular to the dimer plane. Edge states are identified as modes that are localized near the lower and upper edges of the strip, shown in red and blue, respectively. (c),(d)  Spectrum of the full model including the $S=0$ two-triplon bound state. (c) Hybridization between the triplon modes and the bound state, indicated by the color gradient. (d) Identification of edge states. 
\label{fig:M3}}
\end{figure}

 Another signature of the topological nature of bulk bands is the occurrence of protected edge modes. In the following, we consider a strip with $2n$ rows of dimers along the $y$ direction [see Fig.~\ref{fig:M3}(a)] and calculate the excitation spectrum of our model Hamiltonian as a function of the momentum $k_x$ along the infinite direction. As expected from the Chern number calculation, we indeed find edge modes that are localized in the two upper or lower rows of the strip, shown in blue and red, respectively. However, over a large momentum range, the edge states are obscured by the quasi continua of bulk bands with quantized momenta $k_y(j)=\pi j/n$. To analyze this in more detail, we first switch off the bound state [Fig.~\ref{fig:M3}(b)]. In this case one expects\cite{Romhanyi2015} edge states that live between the continua of the upper and lower triplon bands with Chern numbers $C=+2$ and
$C=-2$, respectively. Unfortunately, the gap between the topological Chern bands is almost completely filled by the topologically trivial central triplon bands. As a result, edge states are visible only at momenta 
close to the zone boundary. As a next step, we include the bound state, which hybridizes with the triplon bands and increases the number of topologically non-trivial bands (see Fig.~\ref{fig:Chern}). Edge states remain visible near the zone boundary and show complex evolution as a function of field (Supplementary Information). 


The identification of mode X with an $S=0$ bound state provides a consistent and well-motivated explanation for the experimental data. However, we now consider the possibility that mode X is, instead, a phonon mode. It is unlikely to be an acoustic mode because no intensity is present below the lowest single triplon band and the latter mode shows no hybridization gap. This leaves the possibility that the mode X is an optical phonon. One would expect a phonon mode to be present even when the magnetic form factor suppresses the magnetic intensity or when the temperature washes out the triplon signal. However, the mode X intensity over the range of $\vert\boldsymbol{Q}\vert$ measured in our experiment is nontrivial and suggestive of being present only through the coupling to the triplon modes. Indeed, at $15$ K when the single triplon intensity is absent, mode X is also not present. It is therefore likely that mode X is visible in our experiment only because it is either entirely magnetic or because of a magnetoelastic coupling. It could be that mode X intensity is present at higher $\vert\boldsymbol{Q}\vert$ where the magnetic form factor is small. A consideration of the magnetoelastic coupling suggests that it would take the form of a coupling to a triplon density whereas mode X appears to hybridize linearly with the single triplons which is accounted for naturally in the bound state scenario. Finally, we turn to  observations of low energy phonons coming from other experiments. Previous neutron scattering experiments\cite{Haravifard2012} have only been able to resolve optical phonon modes at energies above $15$ meV while Raman spectra have been interpreted as showing only magnetic modes below about $7$ meV.\cite{Lemmens2000,Gozar2005b}   While these considerations suggest that the new mode X is entirely magnetic and not structural we cannot rule out the phonon scenario and further experiments to look for low energy optical phonons are desirable. 


Using inelastic neutron scattering, we have explored the evolution of the magnetic excitations in the gapped dimer system \scbo\ as a function of a small magnetic field. In addition to the weakly dispersive single-triplon excitations, we identified an $S=0$ two-triplon bound state. This singlet mode is comparatively dispersive and visible in the magnetic structure factor only because it hybridizes with the triplon bands. So far hints of low-energy bound states have been found only in Raman\cite{Gozar2005b,Lemmens2000} and ESR\cite{Nojiri2003} experiments. The presence of low-lying bound states may be responsible for the remarkable sensitivity of triplon coherence to thermal fluctuations as suggested in Ref.~\onlinecite{Honecker2016}. Using the unprecedented insight into the nature of the magnetic couplings in this material, we have determined a minimal spin model that provides an excellent description of the magnetic excitations, including the singlet bound state. For this comprehensive   
model we have calculated the Chern numbers of the bands and the edge-state spectrum. Although the magnetic excitations are much richer than originally proposed, our work shows that \scbo\ is one of the first clear-cut examples of a bosonic topological insulator. For the future, it would be very interesting to learn how to probe the magnetic edge states in this material and how to manipulate them. 
\color{black}


\newpage

\section{Bond Operators}

\noindent
Following Sachdev and Bhatt,\cite{Sachdev1990} we introduce operators that create singlet and triplet states from the vacuum 
\begin{subequations}
\begin{align}
& \vert s \rangle  = s^{\dagger} \vert {\rm VAC} \rangle = \frac{1}{\sqrt{2}} \left( \vert \uparrow\downarrow \rangle - \vert \downarrow\uparrow \rangle    \right), \\
& \vert t_x \rangle  = t_x^{\dagger} \vert {\rm VAC} \rangle = - \frac{1}{\sqrt{2}} \left( \vert \uparrow\uparrow \rangle - \vert \downarrow\downarrow \rangle    \right), \\
& \vert t_y \rangle  = t_y^{\dagger} \vert {\rm VAC} \rangle = \frac{i}{\sqrt{2}} \left( \vert \uparrow\uparrow \rangle + \vert \downarrow\downarrow \rangle    \right), \\
& \vert t_z \rangle  = t_z^{\dagger} \vert {\rm VAC} \rangle = \frac{1}{\sqrt{2}} \left( \vert \uparrow\downarrow \rangle + \vert \downarrow\uparrow \rangle    \right).
\end{align}
\end{subequations}
Together with the constraint 
\begin{equation} s^\dagger s + \sum_{\alpha} t_{\alpha}^{\dagger}t_{\alpha}=1, 
\label{eq:constraint}
\end{equation} 
these operators allow one to navigate in the space of states of two spins one-half. The spin operators themselves are
\begin{subequations}
\begin{align}
S_1^{\alpha} & = \frac{1}{2} \left( s^{\dagger} t_{\alpha} + t_{\alpha}^{\dagger}s - i\epsilon_{\alpha\beta\gamma} t_{\beta}^{\dagger} t_{\gamma}  \right), \\
S_2^{\alpha} & = \frac{1}{2} \left( - s^{\dagger} t_{\alpha} - t_{\alpha}^{\dagger}s - i\epsilon_{\alpha\beta\gamma} t_{\beta}^{\dagger} t_{\gamma}  \right). 
\label{eq:bondoperator}
\end{align}
\end{subequations}
The exchange coupling $\boldsymbol{S}_1 \cdot \boldsymbol{S}_2$ can be written in terms of the bond operators as follows:
\begin{equation}  
-\frac{3}{4} s^{\dagger} s + \frac{1}{4} \sum_{\alpha} t_{\alpha}^{\dagger}t_{\alpha},
\end{equation}
after imposing the constraint.

\section{Lattice Convention and Exchange Couplings}
\label{sec:latt}

Below $395$ K, \scbo\ adopts the tetragonal structure $I\bar{4}2m$ (number $121$). The magnetic ions are the copper Cu$^{2+}$ ions which occupy the $8i$ Wyckoff positions with $x=0.114$ and $z=0.288$. The lattice constants of the tetragonal cell are $a=8.99$~\AA\ and $c=6.648$~\AA. They form a layered structure stacked along the $c$ direction.One such layer is shown in Fig.~1(a) of the main paper. Although there exists a small buckling, the copper bonds A and B are almost coplanar.  The short bonds (marked in blue) between neighboring copper ions are associated with the strongest exchange coupling $J$ which is antiferromagnetic. In the tetragonal primitive cell, the basis in units of the edge length in each direction is 
\begin{subequations}
\begin{align}
A1 : & \left( x,x,z \right), \\
A2 : & \left( -x,-x,z \right), \\
B1 : & \left( x,-x,-z \right), \\
B2 : & \left( -x,x,-z \right).
\end{align}
\end{subequations}
relative both to $(0,0,0)$ and to the body centre $(1/2,1/2,1/2)$. The vertical separation between copper bonds A and B is $(4z-1)c/2$ and the layers are separated by a layer of strontium. 

The point group $D_{2d}$ associated with $I\bar{4}2m$ consists of a $C_2$ rotation around the $c$ axis and $C_2$ rotations around the $a$ axes. Also, there are reflection planes: $[110]$ and $[1\bar{1}0]$ and an $S_4$ with $c$ as the rotation axis. These symmetries constrain the types of exchange that can arise between the copper ions. For example, within nearest neighbor copper bonds labelled by $A$, the possible types of exchange are $S_1^{\rm x} S_2^{\rm x}+S_1^{\rm y} S_2^{\rm y}$, $S_1^{\rm z} S_2^{\rm z}$, $S_1^{\rm x} S_2^{\rm y}+S_1^{\rm y} S_2^{\rm x}$ and $S_1^{\rm x} S_2^{\rm z}+S_1^{\rm y} S_2^{\rm z} - S_1^{\rm z} S_2^{\rm x} - S_1^{\rm z} S_2^{\rm y}$. The latter coupling is an intra-dimer Dzyaloshinskii-Moriya (DM) coupling $\boldsymbol{D}_{ij}\cdot (\boldsymbol{S}_i \times \boldsymbol{S}_j)$. Symmetry also constrains the exchange on the B bonds. Referring to Fig.~1(b) (main paper)  for the lattice directions, the DM vector for the A bonds points in the vertical direction $(0,D,0)$ and for the B bonds in the negative horizontal direction $(-D,0,0)$ where the orientation $i\rightarrow j$ is always from site $1$ to site $2$. 

For Cu$^{2+}$  ($d^9$) with spin one-half, the {\it a priori} superexchange should be mainly isotropic with a smaller DM coupling to leading order in the (weak) spin-orbit coupling with the symmetric anisotropic exchange being weaker still. The nearest neighbor magneto-static dipolar coupling is $0.0088$ meV.

Symmetry does not constrain the exchange at all on any given bond connecting neighboring A and B sites. However, once the exchange is fixed on such a bond, all other such bonds are determined. So there are nine distinct exchange couplings between neighboring dimer bonds. Of these, isotropic exchange $J'$  is expected to be largest followed by the antisymmetric exchange which contributes three parameters to the exchange Hamiltonian. 

We may fix our conventions for the Dzyaloshinskii-Moriya $\boldsymbol{D'}$ vector on the lower right bond oriented from the B1 site to the A2  site in Fig.~1(b) (main paper). We denote the staggered  $x$ component of the DM vector $D'_{\rm s}$ (short blue arrow) and the vertical component $D'_{\parallel}$  (long red arrow). The symmetry operations of the crystal can be used to determine the $\boldsymbol{D'}$  vectors on the other bonds connecting dimers as shown in the figure. All components perpendicular to the plane, $D'_{\perp}$, point upwards out of the page.

We use the following Fourier transform convention throughout
\begin{align}
t^\dagger_{\boldsymbol{i} s\alpha} =\frac{1}{ \sqrt{N}} \sum_{\boldsymbol{k}} t^\dagger_{\boldsymbol{k} s\alpha} \exp\left[ i \boldsymbol{k}\cdot (\boldsymbol{R}_i + \boldsymbol{r}_s) \right],
\end{align}
where $i$ is the Bravais lattice site label with lattice vector $\boldsymbol{R}_i$ and $s$ is the $A$ or $B$ sublattice with two-dimensional lattice vectors $\boldsymbol{r}_A=(0,0)$ and $\boldsymbol{r}_B=a(1/2,1/2)$.

\section{Interactions in Bond Operator Language}

\noindent
Using the mapping to bond operators, Eq.~\ref{eq:bondoperator} and imposing the constraint Eq.~\ref{eq:constraint} we find
\begin{equation}
H_{\rm Intra-DM} =  \frac{iD}{2} \sum_{i\in A} \left( s^\dagger t_y - t_y^\dagger s \right) -  \frac{iD}{2} \sum_{i\in B} \left( s^\dagger t_x - t_x^\dagger s \right).
\end{equation}
We may remove these terms, thus diagonalizing the intra-dimer Hamiltonian, by performing a unitary transformation on the singlet and triplet operators. Instead of using the full transformation, we note that $D/J \sim 0.1$ so we perform the rotation to order $D/J$.  In particular, on the $A$ bonds, we go to $\tilde{s}^\dagger$ and $\tilde{t}_{\alpha}^\dagger$ operators which are
\begin{equation}
\left( \begin{array}{c}  \tilde{s}^\dagger \\ \tilde{t}_{x}^\dagger  \\ \tilde{t}_{y}^\dagger  \\ \tilde{t}_{z}^\dagger  \end{array} \right) = \left( \begin{array}{cccc} 1 & 0 & i\frac{D}{2J} & 0 \\  0 & 1 & 0 & 0 \\  i\frac{D}{2J} & 0 & 1 & 0 \\  0 & 0 & 0 & 1   \end{array} \right) \left( \begin{array}{c}  s^\dagger \\  t_{x}^\dagger  \\ t_{y}^\dagger  \\ t_{z}^\dagger  \end{array} \right) 
\label{eq:DMA}
\end{equation}
while on the $B$ bonds we have
\begin{equation}
\left( \begin{array}{c}  \tilde{s}^\dagger \\ \tilde{t}_{x}^\dagger  \\ \tilde{t}_{y}^\dagger  \\ \tilde{t}_{z}^\dagger  \end{array} \right) = \left( \begin{array}{cccc} 1 & -i\frac{D}{2J} & 0 & 0 \\  -i\frac{D}{2J} & 1 & 0 & 0 \\  0 & 0 & 1 & 0 \\  0 & 0 & 0 & 1   \end{array} \right) \left( \begin{array}{c}  s^\dagger \\  t_{x}^\dagger  \\ t_{y}^\dagger  \\ t_{z}^\dagger  \end{array} \right).
\label{eq:DMB}
\end{equation}

Now we consider exchange coupling neighboring dimers. The most important of the symmetry allowed exchange couplings is isotropic with coupling $J'$. We write this in terms of triplet operators on each $J'$ bond and sum over the two bonds connecting each pair of dimers.  This gives $H_3 + H_4$ where
\begin{align}
H_3 & = i\frac{J'}{2} \left( - \right)^{s(j)} \sum_{\langle ia,jb \rangle,\alpha} \epsilon_{\alpha\beta\gamma} t_{i\beta}^\dagger t_{i\gamma} \left( t_{j\alpha} + t_{j\alpha}^\dagger \right) \label{eq:H3}, \\
H_4 & = - \frac{J'}{2}  \sum_{\langle ia,jb \rangle,\alpha} \left( t_{i\beta}^\dagger t_{j\beta}^\dagger t_{i\gamma} t_{j\gamma} - t_{i\beta}^\dagger t_{j\gamma}^\dagger t_{i\gamma} t_{j\beta} \right).  \label{eq:H4}
\end{align}
Here, $s(j)$ is the site label, $1$ or $2$ of site $j$ and there is no double-counting on bonds in the sums. The three-body terms, $H_3$ may create or annihilate triplets on two of the four dimer bonds neighboring a given dimer. The single particle triplon hopping term originating from $J'$ on its own vanishes owing to the geometrical frustration of the lattice.

The inter-dimer DM interaction discussed in Section~\ref{sec:latt}, when written in terms of singlet and triplet operators, generates hopping terms between A and B dimers. Of the three DM couplings, the contribution coming from $D'_{\rm s}$  cancels leaving hopping depending only on  $D'_{\parallel}$ and $D'_{\perp}$. The hopping Hamiltonian originating from these terms takes the form
\[  H_{{\rm InterDM}}^{(2)} = \sum_{\boldsymbol{r}\in A} \sum_{\boldsymbol{\delta}} \sum_{\mu,\nu=x,y,z} t^{\dagger}_{\boldsymbol{r}+\boldsymbol{\delta},\mu} M_{\mu\nu}^{\rm BA}(\boldsymbol{\delta}) t_{\boldsymbol{r},\nu} + \sum_{\boldsymbol{r}\in B} \sum_{\boldsymbol{\delta}} \sum_{\mu,\nu=x,y,z} t^{\dagger}_{\boldsymbol{r}+\boldsymbol{\delta},\mu} M_{\mu\nu}^{\rm AB}(\boldsymbol{\delta})t_{\boldsymbol{r},\nu},
\]
where 
\[ 
\boldsymbol{M}^{\rm BA}(\pm \boldsymbol{x} ) = \frac{1}{2} \left(  
\begin{array}{ccc}    
0 & -D'_{\perp} & 0 \\
D'_{\perp} & 0 & \pm D'_{\parallel} \\
0 & \mp D'_{\parallel} & 0 
\end{array}   \right)
\]
\[ 
\boldsymbol{M}^{\rm AB}(\pm \boldsymbol{x} ) = \frac{1}{2} \left(  
\begin{array}{ccc}    
0 & D'_{\perp} & 0 \\
-D'_{\perp} & 0 & \pm D'_{\parallel} \\
0 & \mp D'_{\parallel} & 0 
\end{array}   \right)
\]
\[ 
\boldsymbol{M}^{\rm AB}(\pm \boldsymbol{y} ) = \frac{1}{2} \left(  
\begin{array}{ccc}    
0 & D'_{\perp} & \mp D'_{\parallel} \\
-D'_{\perp} & 0 &  0 \\
\pm D'_{\parallel} &  0 & 0 
\end{array}   \right)
\]
\[ 
\boldsymbol{M}^{\rm BA}(\pm \boldsymbol{y} ) = \frac{1}{2} \left(  
\begin{array}{ccc}    
0 & -D'_{\perp} & \mp D'_{\parallel} \\
D'_{\perp} & 0 &  0 \\
\pm D'_{\parallel} &  0 & 0 
\end{array}   \right)
\]

We now discuss the effects of the transformation of the ground state and triplon operators brought about by the presence of the intra-dimer DM interaction  $\boldsymbol{D}$.  Whereas the $J'$ coupling does not lead to hopping on its own, the rotation introduces hopping terms to order $D J'/J$. One may show that these terms amount to a shift of the $D'_{\parallel}$ hopping terms coming from the inter-dimer DM exchange. More precisely, $D'_{\parallel}\rightarrow D'_{\parallel}+DJ'/2J$. 

We have understood that, besides isotropic exchange and DM couplings, there are two further anisotropic exchange couplings within each dimer allowed by symmetry. These terms lead to triplon terms of the form $t_z^\dagger t_z$ and $t_x^\dagger t_y + t_y^\dagger t_x$ for $S_1^z S_2^z$ and $S_1^x S_2^y+S_1^y S_2^x$ respectively. For the inter-dimer exchange, we have considered four couplings out of a possible nine one of which - the isotropic exchange coupling - does not contribute to the hopping to lowest order in $J'$. The remaining anisotropic terms are symmetric and hence do not lead to leading order hopping terms.

The authors of Ref.~\onlinecite{Romhanyi2015} observed that the single triplon Hamiltonian with $J$, the DM couplings and hopping between second neighbor dimers has equivalent sublattices A and B. This means that the Brillouin zone can be unfolded so that there are $3$ bands in the zone. The explicit transformation that makes this translational invariance manifest is given in the supplementary material of Ref.~\onlinecite{Romhanyi2015}. We note that this translational symmetry is broken by the presence of the pseudo-dipolar coupling $J_{\rm xy}$.

\section{Perturbation Theory in $J'$}
\label{sec:PT}

Here we outline the perturbation theory in $J'$ within a fixed triplet number sector and in the absence of DM interactions. To zeroth order in perturbation theory, the state of the system is determined by $J$ alone - it is a direct product of singlets in the ground state and degenerate triplet states on each site. We organize the perturbation theory as an effective Hamiltonian of the form $H_{\rm eff}=H^{(0)} + (J'/J) H^{(1)} + (J'/J)^2 H^{(2)}+\ldots$ computed from the full Hamiltonian $H=H^{(0)}  + V$. To do this we introduce a projector $\mathcal{P}^{(N)}$ onto the degenerate zeroth order triplet sector $\mathcal{M}^{(N)}$ with $N$ triplets and a resolvent operator $\mathcal{R}^{(N)} = \sum_{E \notin \mathcal{M}} \vert E \rangle\langle E \vert / (\epsilon_{G} - \epsilon_{E})$ that connects the system to states $\vert E \rangle$ outside this sector. Here $\epsilon_{S}$ is the energy computed from $H^{(0)}$ for state $S$. Then $H_{\rm eff}=H^{(0)} + \mathcal{P}^{(N)}V\mathcal{P}^{(N)} + \mathcal{P}^{(N)}V\mathcal{R}^{(N)}V\mathcal{P}^{(N)} + \ldots$.

By carrying out perturbation theory in powers of $J'/J$ we see that triplet hopping first appears to order $(J'/J)^6$.  So, the triplon modes acquire only a weak dispersion in $J'$. The ground state is even more resistant to the presence of $J'$: remarkably, in the $J$, $J'$ model, the direct product state of singlets on the dimer bonds is the exact ground state for $J'<J'_c \approx 0.68J$. This is the celebrated result of Shastry and Sutherland.\cite{Shastry1981,Miyahara1999}

Although hopping of triplons is suppressed up to the order $(J'/J)^6$, the energy of a single triplon is shifted by the presence of the inter-dimer exchange even at the second order. We find, 
\begin{equation}  \Delta \equiv J -  \frac{J'^2}{J} - \frac{J'^3}{2J^2} + \ldots 
\label{eq:Delta}
\end{equation}

We further develop this perturbation theory below to address the presence and nature of the bound states.

\section{Single Triplon Hopping}
\label{sec:ST}

The Hamiltonian including intra-dimer isotropic and anisotropic exchange contributions as well as inter-dimer DM is given by  $H_{\rm ST}= \int_{\boldsymbol{k}} T_{\boldsymbol{k}}^\dagger \Lambda_{\boldsymbol{k}} T_{\boldsymbol{k}}$ where  $T_{\boldsymbol{k}}^\dagger=(\boldsymbol{t}_{\boldsymbol{k},A}^\dagger,\boldsymbol{t}_{\boldsymbol{k},B}^\dagger)$ with $\boldsymbol{t}_{\boldsymbol{k},\alpha}^\dagger = (t_{\boldsymbol{k},\alpha,x}^\dagger,t_{\boldsymbol{k},\alpha,y}^\dagger,t_{\boldsymbol{k},\alpha,z}^\dagger)$, and 
\[  \Lambda_{\boldsymbol{k}}  =  \left(  \begin{array}{cc} A_{\boldsymbol{k}} & C_{\boldsymbol{k}} \\ C^\dagger_{\boldsymbol{k}} & B_{\boldsymbol{k}}  \end{array} \right).   \]
 The coupling matrix on sublattice $A$ is given by
\[
A_{\boldsymbol{k}} =  \left(  
\begin{array}{ccc}    
J & ig_z \mu_{\rm B} B + \frac{J_{\rm xy}}{2} & 0  \\
-ig_z \mu_{\rm B} B + \frac{J_{\rm xy}}{2} & J & 0 \\
0 & 0 & J + \frac{J_{\rm zz}}{2}  \end{array}  \right),
\]
while the corresponding matrix $B_{\boldsymbol{k}}$ on sublattice $B$ is obtained from $A_{\boldsymbol{k}}$ by inverting the sign of $J_{\rm xy}$. Finally, the coupling between the sublattices takes the form
\[
C_{\boldsymbol{k}} =  \left(  
\begin{array}{ccc}  0 & 2D'_{\perp}\cos \frac{k_x}{2}\cos \frac{k_y}{2} & -iD'_{\parallel} \sin \frac{k_x - k_y}{2} \\
-2D'_{\perp}\cos \frac{k_x}{2}\cos \frac{k_y}{2} & 0 & -iD'_{\parallel} \sin \frac{k_x + k_y}{2} \\
iD'_{\parallel} \sin \frac{k_x - k_y}{2} & iD'_{\parallel} \sin \frac{k_x + k_y}{2} & 0 
 \end{array}  \right),
\]
where for brevity,  we have absorbed the shift $D'_{\parallel}\rightarrow D'_{\parallel}+DJ'/2J$ in a re-definition of the coupling constant. Following previous work,\cite{Romhanyi2015}  
we also include a hopping between next-neighbor dimers with coupling $J_{\rm FN}$ which enters into the diagonal elements as $J\rightarrow J + 2J_{\rm FN}\cos(\frac{k_x + k_y}{2})\cos(\frac{k_x - k_y}{2})$.

The qualitative effects of each exchange term on the dispersion are as follows. The $J$ term simply gives a gap to the triplons. The inter-dimer DM couplings break the degeneracy of the triplon bands over most of the Brillouin zone. They leave the central mode dispersionless and the dispersion of the upper and lower bands symmetrical about the middle band so that the energies do not change under sign flips of the DM couplings. The $D'_{\parallel}$ coupling leads to linear touching at the Brillouin centre and corners. $D'_{\perp}$  leads to quadratic touching at the Brillouin zone corners. In order to break the symmetry between the upper and lower bands and to introduce a dispersion for the central mode we introduce $K_{\rm hop}$. The modes measured in \scbo\ appear to be asymmetrical in the above sense.

The intra-dimer anisotropic exchange $J_{\rm zz}$ breaks any degeneracy between one pair of mode and the other two pairs. The intra-dimer coupling $J_{\rm xy}$ completely lifts the degeneracy of the modes except at the Brillouin zone corners. This implies that there can be no transformation that allows one to render sublattices $A$ and $B$ equivalent.

\section{Anomalous Terms}
\label{sec:anomalous}

\begin{figure}[htpb]
\includegraphics[width=0.6\columnwidth]{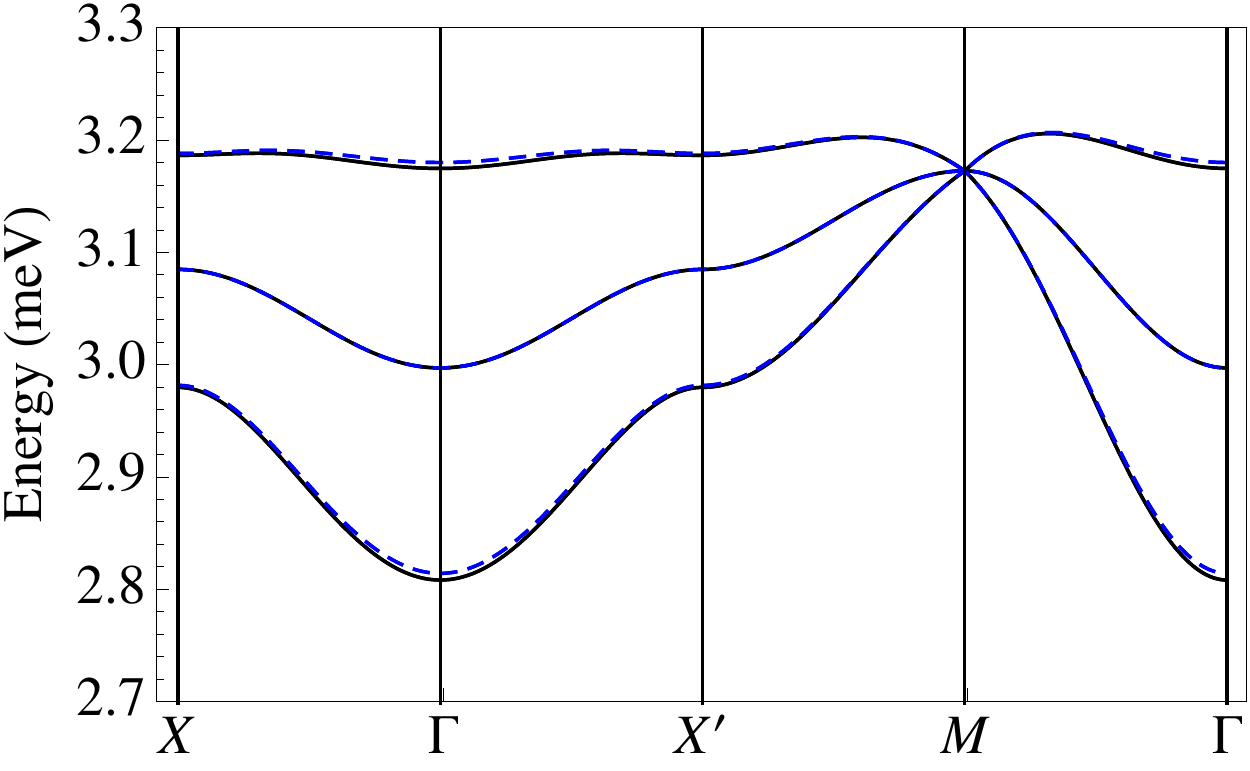}
\caption{
{\bf Effect of the anomalous terms.} Comparison of the triplon excitation spectra with (dashed lines) and without (solid lines) the anomalous terms, using the model of Ref.~[\onlinecite{Romhanyi2015}] 
in zero field. 
\label{fig:anomalous}}
\end{figure}

Anomalous quadratic terms $t_{A\alpha}^\dagger t_{B\beta}^\dagger$ and $t_{A\alpha} t_{B\beta}$ arise from the bond-operator representation of the coupling between spins of neighboring dimers
and are proportional to $D'$ and $DJ'/J$, exactly as the triplon hopping terms. Including these terms, the quadratic triplon Hamiltonian can be written as
\[  \tilde{H}_{\rm ST}=\frac 12\int_{\boldsymbol{k}} \left(T_{\boldsymbol{k}}^\dagger,T_{-\boldsymbol{k}}  \right)    \left(  \begin{array}{cc}  \Lambda_{\boldsymbol{k}} & \Pi_{\boldsymbol{k}} \\ \Pi^\dagger_{\boldsymbol{k}} & \Lambda^t_{-\boldsymbol{k}}  \end{array} \right) \left(  \begin{array}{c} T_{\boldsymbol{k}} \\ T_{-\boldsymbol{k}}^\dagger   \end{array} \right),  \]
where the six-dimensional triplon operators $T_{\boldsymbol{k}}$ and coupling matrices $\Lambda_{\boldsymbol{k}}$ have been defined in Sec.~\ref{sec:ST}. The coupling matrix for the 
anomalous terms is given by 
\[  \Pi_{\boldsymbol{k}}  =  \left(  \begin{array}{cc} 0 & C_{\boldsymbol{k}} \\ C^\dagger_{\boldsymbol{k}} & 0  \end{array} \right),   \]
and is identical to the triplon hopping matrix between neighboring dimers. It has been argued by Romhanyi {\it et al.}\cite{Romhanyi2015} that the anomalous terms have a negligible effect since they do not change the triplon energies to linear order (they renormalize the single-triplon parameters to second order in perturbation theory). In order to investigate this further, we compare the triplon excitation 
spectra in the presence and absence of the anomalous terms. It is straightforward to diagonalize the quadratic triplon Hamiltonian $\tilde{H}_{\rm ST}$ by a Bogoliubov transformation. This includes the anomalous terms to infinite order. In Fig.~\ref{fig:anomalous} the triplon excitation spectra are shown in zero field and for the parameters of Ref.~[\onlinecite{Romhanyi2015}]. The corrections due to the anomalous terms are indeed very small and affect only the upper and lower triplon bands away from the $M$ point. 

Treating the anomalous terms in second order perturbation theory it is possible to work out the leading terms that renormalize the triplon dispersion. We obtain triplon hopping terms $\sim \frac{(D'_\perp)^2}{J} (t_{ix}^\dagger t_{jx}+t_{iy}^\dagger t_{jy}+\textrm{h.c.})$ and $\sim \frac{(D'_\parallel)^2}{J} (t_{ix}^\dagger t_{jy}+t_{iy}^\dagger t_{jx}+\textrm{h.c.})$ between NNN dimers, which are on the same sub-lattice.   Note that we have already included an isotropic NNN hopping term $J_{\rm FN}$ in the single-triplon Hamiltonian. Similar hopping terms are also generated between 3rd nearest neighbor dimers. 
Relative to the triplon bandwidth $W\sim D'$ we therefore expect corrections of the order of $D'/J\simeq 3 \%$.

\section{Singlet Bound State}
\label{sec:SBS}

\begin{figure}[htpb]
\centering
\hspace{3cm}
\centering
\includegraphics[width=0.95\columnwidth]{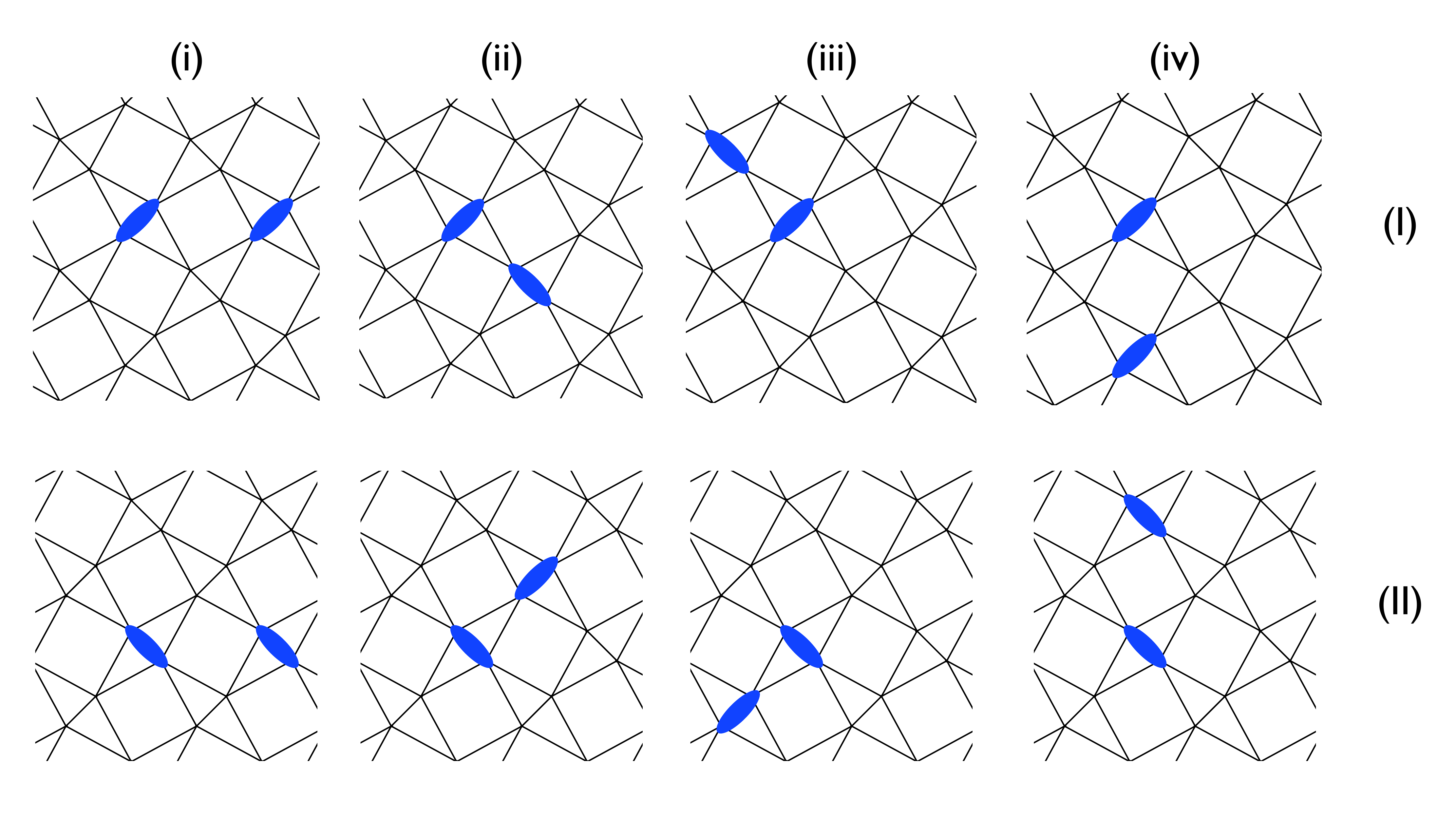}
\caption{ 
{\bf Basis states of two triplon bound states.} States are divided into two sectors (I and II) which form different blocks in the Hamiltonian and each sector contains four states (i)-(iv).
\label{fig:S2}}
\end{figure}

We consider two triplon bound states as candidates for the new dispersing feature in the neutron scattering data. We expect the lowest energy bound state to belong to the total $S=0$ sector with wavefunction
\[  \vert \Phi_{\boldsymbol{K}}  \rangle = \frac{1}{\cal{N}} \sum_{\boldsymbol{q},\alpha} \Phi_{\boldsymbol{K},\boldsymbol{q}}  t^\dagger_{\frac{\boldsymbol{K}}{2}+\boldsymbol{q},\alpha} t^\dagger_{\frac{\boldsymbol{K}}{2}-\boldsymbol{q},\alpha}  \vert 0\rangle.  \]
In general, the bound state will have four modes corresponding to the four choices of sublattice pairs. The bound state can be captured within a perturbative calculation even to low order as the $J'$ term induces a correlated hopping of neighboring triplons which lowers the energy relative to the two-triplon continuum.  To see this, we consider the two sets of four states shown in Fig.~\ref{fig:S2}.

In momentum space, we define the bound state creation and annihilation operators by
\begin{align*}
\Phi_{\boldsymbol{K}}^{(Ii)\dagger} & = \frac{1}{\sqrt{N}} \sum_{\boldsymbol{q},\alpha}  t^{\dagger}_{\frac{\boldsymbol{K}}{2}+\boldsymbol{q}A\alpha} t^{\dagger}_{\frac{\boldsymbol{K}}{2}-\boldsymbol{q}A\alpha} e^{i \left( \frac{\boldsymbol{K}}{2} - \boldsymbol{q} \right) \cdot\boldsymbol{x}}, \\
\Phi_{\boldsymbol{K}}^{(Iii)\dagger} & = \frac{1}{\sqrt{N}} \sum_{\boldsymbol{q},\alpha}  t^{\dagger}_{\frac{\boldsymbol{K}}{2}+\boldsymbol{q}A\alpha} t^{\dagger}_{\frac{\boldsymbol{K}}{2}-\boldsymbol{q}B\alpha} e^{i \left( \frac{\boldsymbol{K}}{2} - \boldsymbol{q} \right) \cdot \left(\frac{ \boldsymbol{x}-\boldsymbol{y}}{2}\right)}, \\
\Phi_{\boldsymbol{K}}^{(Iiii)\dagger} & = \frac{1}{\sqrt{N}} \sum_{\boldsymbol{q},\alpha}  t^{\dagger}_{\frac{\boldsymbol{K}}{2}+\boldsymbol{q}A\alpha} t^{\dagger}_{\frac{\boldsymbol{K}}{2}-\boldsymbol{q}B\alpha} e^{i \left( \frac{\boldsymbol{K}}{2} - \boldsymbol{q} \right) \cdot \left(\frac{ \boldsymbol{y}-\boldsymbol{x}}{2}\right)}, \\
\Phi_{\boldsymbol{K}}^{(Iiv)\dagger} & = \frac{1}{\sqrt{N}} \sum_{\boldsymbol{q},\alpha}  t^{\dagger}_{\frac{\boldsymbol{K}}{2}+\boldsymbol{q}A\alpha} t^{\dagger}_{\frac{\boldsymbol{K}}{2}-\boldsymbol{q}A\alpha} e^{-i \left( \frac{\boldsymbol{K}}{2} - \boldsymbol{q} \right) \cdot\boldsymbol{y}},
\end{align*}
with similar definitions for two triplon states in sector II.

The spectra for the bound states for sector I to third order in perturbation theory are found by diagonalizing the matrix\cite{Totsuka2001}
\[ 
H_{\rm Bound}^{(S-I)} = \left(  
\begin{array}{cccc}    
2\Delta + V_{\rm NNN} & J_{\rm NN} & J_{\rm NN}e^{-i k_x} & 0 \\
J_{\rm NN} & 2\Delta + V_{\rm NN} & J_{3{\rm rd}} & -J_{\rm NN} \\
J_{\rm NN}e^{i k_x} & J_{3{\rm rd}} & 2\Delta + V_{\rm NN} & -J_{\rm NN}e^{-i k_y}  \\
0 & -J_{\rm NN}& - J_{\rm NN}e^{i k_y}   & 2\Delta + V_{\rm NNN} 
\end{array}   \right),
\]
where the potentials and hopping terms are
\begin{align*}
V_{\rm NN} & =  -J'  + \frac{J^{'2} }{2J} + \frac{J^{'3}}{J^2}, \\
V_{\rm NNN} & = - \frac{J^{'3}}{2J^2}, \\
J_{\rm NN} & = - \frac{J^{'2} }{2J} - \frac{J^{'3}}{4J^2}, \\
J_{3{\rm rd}} & = - \frac{J^{'2} }{2J}.
\end{align*}
For sector II, the Hamiltonian is 
\[ 
H_{\rm Bound}^{(S-II)} = \left(  
\begin{array}{cccc}    
2\Delta + V_{\rm NNN} & J_{\rm NN} & J_{\rm NN}e^{-i k_x} & 0 \\
J_{\rm NN} & 2\Delta + V_{\rm NN} & J_{3{\rm rd}} & -J_{\rm NN} \\
J_{\rm NN}e^{i k_x} & J_{3{\rm rd}} & 2\Delta + V_{\rm NN} & -J_{\rm NN}e^{i k_y}  \\
0 & -J_{\rm NN}& - J_{\rm NN}e^{-i k_y}  & 2\Delta + V_{\rm NNN}
\end{array}   \right).
\]
So the dispersion for sector II is related to that for sector I by a reflection about $k_x =0$.

\section{Hybridization}
\label{sec:hyb}

The DM interactions supply a means for the singlet bound states and single triplon modes to hybridize. The three-body triplon Hamiltonian arising from $J'$ (Eq.~\ref{eq:H3}) directly couples the two. A second contribution arises from Eq.~\ref{eq:H4} after performing the transformations Eq.~\ref{eq:DMA} and \ref{eq:DMB} on a single triplon operator. This, second, contribution merely renormalizes the first. 
The hybridization matrix between the six single triplon states and the four two-triplon singlet states in sector I is
\begin{equation}
\left( \begin{array}{cccc}   
0 & -i\frac{D'_s}{\sqrt{3}} &  -i\frac{D'_s}{\sqrt{3}} & 0 \\
0 &  0 &  0 &  0 \\
0 &  0 &  0 &  0 \\
0 &  0 &  0 &  0 \\
0 & -i\frac{D'_\parallel}{\sqrt{3}} e^{i\boldsymbol{k}\cdot\left( \frac{\boldsymbol{x}-\boldsymbol{y}}{2} \right) } &  -i\frac{D'_\parallel}{\sqrt{3}}e^{-i\boldsymbol{k}\cdot\left( \frac{\boldsymbol{x}-\boldsymbol{y}}{2} \right) } & 0 \\
0 & -i\frac{D'_\perp}{\sqrt{3}} e^{i\boldsymbol{k}\cdot\left( \frac{\boldsymbol{x}-\boldsymbol{y}}{2} \right) } &  i\frac{D'_\perp}{\sqrt{3}}e^{-i\boldsymbol{k}\cdot\left( \frac{\boldsymbol{x}-\boldsymbol{y}}{2} \right) } & 0 
\end{array} \right),
\end{equation}
where the six single triplon states are organized as in Section~\ref{sec:ST}. As before, we have absorbed the shift of the coupling constants,  $D'_\parallel\rightarrow D'_\parallel  + DJ'/2J$ and $D'_s\rightarrow D'_s  -DJ'/2J$ (arising from the diagonalization of the local dimer Hamiltonian) in a re-definition of parameters. For sector II, we find
\begin{equation}
\left( \begin{array}{cccc}   
0 & i\frac{D'_\parallel}{\sqrt{3}}e^{i\boldsymbol{k}\cdot\left( \boldsymbol{x}+\boldsymbol{y} \right) }  &  i\frac{D'_\parallel}{\sqrt{3}} & 0 \\
0 &  0 &  0 &  0 \\
0 & i\frac{D'_\perp}{\sqrt{3}}e^{i\boldsymbol{k}\cdot\left( \boldsymbol{x}+\boldsymbol{y} \right) }  &  -i\frac{D'_\perp}{\sqrt{3}} & 0 \\
0 &  0 &  0 &  0 \\
0 & i\frac{D'_s}{\sqrt{3}} e^{i\boldsymbol{k}\cdot\left( \frac{\boldsymbol{x}+\boldsymbol{y}}{2} \right) } &  i\frac{D'_s}{\sqrt{3}}e^{i\boldsymbol{k}\cdot\left( \frac{\boldsymbol{x}+\boldsymbol{y}}{2} \right) } & 0 \\
0 &  0 &  0 &  0 
\end{array} \right).
\end{equation}

\section{Dynamical Structure Factor}

\noindent
The inelastic neutron scattering intensity is 
\begin{align}
 I\left(\boldsymbol{K},\omega  \right) & \propto \vert F\left(\boldsymbol{K} \right) \vert^2  P_{\alpha\beta}\left(\boldsymbol{K} \right) \sum_n \langle 0 \vert S^\alpha (\boldsymbol{K}) \vert n\rangle \langle n \vert S^\beta (-\boldsymbol{K}) \vert 0\rangle \delta\left( \omega - \omega_n \left( \boldsymbol{K} \right) \right)
\label{eq:structurefactor}
\end{align}
for scattering wavevector $\boldsymbol{K}$ and energy loss $\omega$. The copper Cu$^{2+}$ form factor is $F\left(\boldsymbol{K} \right)$ and $P_{\alpha\beta}\left(\boldsymbol{K} \right)$ is the transverse projector $\delta_{\alpha\beta}-\hat{K}_\alpha\hat{K}_\beta$. The spin operator here is 
\begin{eqnarray*}
S^\alpha (\boldsymbol{K}) & = &  \frac{1}{\sqrt{N}} \sum_{\boldsymbol{R}_i} e^{i\boldsymbol{K}\cdot\boldsymbol{R}_i } \left\{ 
e^{i\boldsymbol{K}\cdot\boldsymbol{r}_{A1}} S^\alpha_{A1}(\boldsymbol{R}_i) + e^{i\boldsymbol{K}\cdot\boldsymbol{r}_{A2}} S^\alpha_{A2}(\boldsymbol{R}_i) \right.\\
& & \left. + e^{i\boldsymbol{K}\cdot\boldsymbol{r}_{B1}} S^\alpha_{B1}(\boldsymbol{R}_i) +e^{i\boldsymbol{K}\cdot\boldsymbol{r}_{B2}} S^\alpha_{B2}(\boldsymbol{R}_i)  
\right\},
\end{eqnarray*}
where $\boldsymbol{R}_i$ labels the primitive tetragonal lattice vectors and $\boldsymbol{r}_{A1}$ etc. label the basis given in Section~\ref{sec:latt}.

We introduce the bond operator representation and carry out the unitary rotation of the singlet and triplet operators necessitated by the presence of the intra-dimer DM interaction. After acting on the vacuum the result is
\begin{align*}
& S^\alpha (-\boldsymbol{K}) \vert 0\rangle = \left\{
i \sin\left( x( K_x + K_y )\right) t_{A}^{\alpha\dagger}(-\boldsymbol{K}) 
+ i  e^{ih K_z } \sin\left( x( K_x - K_y )\ \right) t_{B}^{\alpha\dagger}(-\boldsymbol{K}) \right. \\
&\left. + \frac{D}{J} \cos\left(  x( K_x + K_y ) \right) \epsilon_{\alpha y \beta} t_{A}^{\beta\dagger}(-\boldsymbol{K}) 
- \frac{D}{J} e^{ih K_z } \cos\left(  x( K_x - K_y )\right) \epsilon_{\alpha x \beta} t_{B}^{\beta\dagger}(-\boldsymbol{K}) 
\right\} \vert 0\rangle,
\end{align*}
where $h= (4z-1)/2$ is the distance along the $c$ direction between dimers nominally in the same layer. Finally, we transform the triplon operators into the basis $\{ | n\rangle\}$  of eigenmodes to calculate the matrix elements entering Eq.~\ref{eq:structurefactor}.

\section{Berry Curvature, Thermal Hall Effect and Chern Number}

The Berry curvature of a band in a crystalline medium is a fictitious local magnetic field that depends on the Bloch wavefunction $\vert \psi^{(n)} \left( \boldsymbol{k} \right) \rangle$  of the band. The analogue vector potential or Berry connection is $A^{(n)}_{\mu}\left( \boldsymbol{k} \right)=\langle \psi^{(n)} \left( \boldsymbol{k} \right) \vert \partial_{\mu}  \vert \psi^{(n)} \left( \boldsymbol{k} \right) \rangle$ from which the Berry curvature may be found from $F^{(n)}_{\mu\nu}=\partial_{\mu}A^{(n)}_{\nu}-\partial_{\nu}A^{(n)}_{\mu}$ - the derivatives being taken in crystal momentum space. We computed the Berry curvature using the link variable method of Ref.~\onlinecite{Fukui2005}. The integral of the Berry curvature $F_{xy}$ over a 2D Brillouin zone is an integer topological invariant - the first Chern number - which simply measures the number of times the mapping $F$ from the Brillouin zone torus covers a torus
\[
C_n = \frac{1}{2\pi i}\int_{\rm BZ} d^2\boldsymbol{k} F^{(n)}_{xy}\left( \boldsymbol{k} \right).
\]
The sum of Chern numbers of all bands is zero and $C_n$ itself is zero unless time reversal $H(\boldsymbol{k})=H^{*}(-\boldsymbol{k})$ is broken. 
This invariant is well-defined only when the band index is well-defined so the bands should not have touching points. Such touching points exist in our model at the corner of the Brillouin zone, irrespective of the field $B$ perpendicular to the plane. In the calculation we include a small transverse field component to lift this degeneracy. It is well-known that, when a pair of bands do touch and separate, the net Chern number of the pair remains unchanged. Since the total Chern number of the pair turns out to be non-zero, the topology remains protected even in the limit of zero transverse field. 
The Chern number is connected to observable quantities. In integer quantum Hall systems, for example, the Chern number is related to the quantized Hall conductance.\cite{Thouless1982} A bosonic analogue of this result for the case when the band is thermally populated is the thermal Hall effect which, rather than depending on the Chern number and yielding a quantized response instead depends on the magnitude of the Berry curvature in thermally occupied parts of the band.  

\begin{figure}[htpb]
\includegraphics[width=0.8\columnwidth]{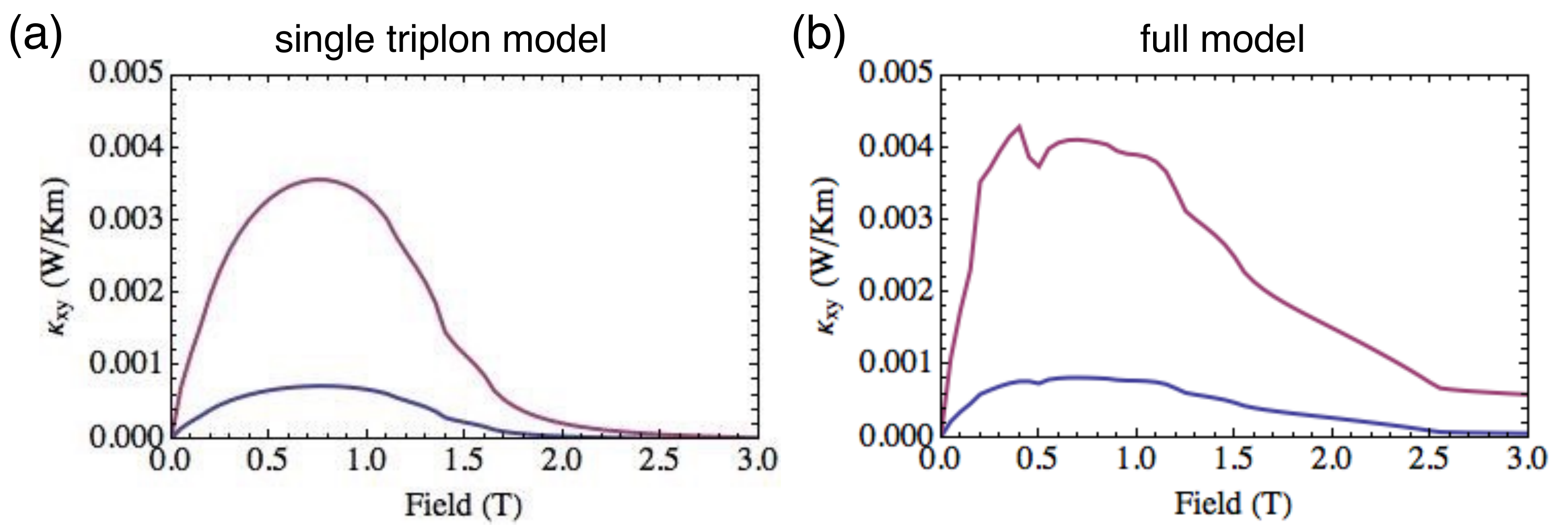} 
\caption{{\bf Calculated thermal Hall conductivity in \scbo.} Plot of $\kappa_{xy}$ as a function of the magnetic field applied along the $c$ direction at $5$ K (blue) and $7.5$ K (red). 
\label{fig:THE}}
\end{figure}

We assume that the transverse thermal conductivity comes from the response of triplons to the Berry curvature in the triplon bands as given by
\begin{equation}  \kappa_{\alpha\beta} = -\frac{k_{\rm B}^2 T}{\hbar L} \sum_{\mathbf{k},n} c_2[ \rho (\omega_{n,\mathbf{k}}) ] F^{(n)}_{\alpha\beta}(\mathbf{k}) 
\label{eqn:TMHE}
\end{equation}
where $F^{(n)}_{\alpha\beta}(\mathbf{k})$ is the Berry curvature in the $n$th band and 
\[  c_2 (\rho) \equiv (1+\rho) \left( \log \frac{1+\rho}{\rho}  \right)^2 - \left( \log \rho \right)^2 - 2{\rm Li}_2 (-\rho).  \]
Here $\rho=(\exp(\beta\omega)-1)^{-1}$ is the Bose distribution function and ${\rm Li}$ is the polylog function. $L=3.32$\AA\ is the layer thickness. The formula above was derived for magnons by Matsumoto and Murakami\cite{Matsumoto2011} building on work by Katsura, Nagaosa and Lee.\cite{Katsura2010} In the latter work, the magnon rotational motion was omitted leading to a formula that depends only on the Berry curvature in the low temperature limit. The formula is directly applicable to the triplon problem. Fig.~\ref{fig:THE} shows the thermal Hall conductivity $\kappa_{xy}$ over a range of magnetic fields at $5$ K and $7.5$ K in \scbo\ including only the single triplon sector. The maximum at around $0.7$ T is due to the maximum in the low energy density of triplon states at this field owing to the fact that the touching point see-saws from the Brillouin zone corner at $0$ T to the zone centre at $1.4$ T with the average energy anchored by the $S=0$ triplon. The topological transition is visible as a kink in the Hall effect resulting from a spike in the Berry curvature at the touching point in the Brillouin zone.

The result of the Hall effect calculation in the presence of our model including the singlet bound state using parameters obtained by fitting the INS data is shown in Fig.~\ref{fig:THE}(b). While the Hall effect again exhibits a broad maximum as the field is increased, the detailed variation is quite different from the case with only single triplons. This is largely due to the fact that the hybridization with the bound state increases the number of topological transitions as a function of field. 

To appreciate this we calculated the Chern numbers of the bands in a very small transverse field that gaps out the Brillouin zone corners so that the Chern number calculation is straightforwardly well-defined. The main text has a figure showing the Chern numbers of the low energy bands at four fields. These figure demonstrates that the possible Chern numbers and number of topologically nontrivial bands is increased relative to the single triplon case. In addition, nontrivial bands persist well above the $1.4$ T upper critical field of the single triplon case. The kinks in the thermal Hall effect correlate with topological transitions between pairs of bands with the effect being greatest at about $0.6$ T and again at $2.6$ T where the Chern number of the lowest-lying nontrivial band changes. Further transitions in higher bands occur at a succession of fields between $0$ T and $3$ T.

It is evident from Fig.~\ref{fig:THE} that the thermal Hall effect does not imply the existence of Chern bands. The robust observable consequence of Chern bands is the presence of chiral edge states as we now discuss.

\section{Field Dependence of the Edge States}

\begin{figure}[htpb]
\includegraphics[width=\columnwidth]{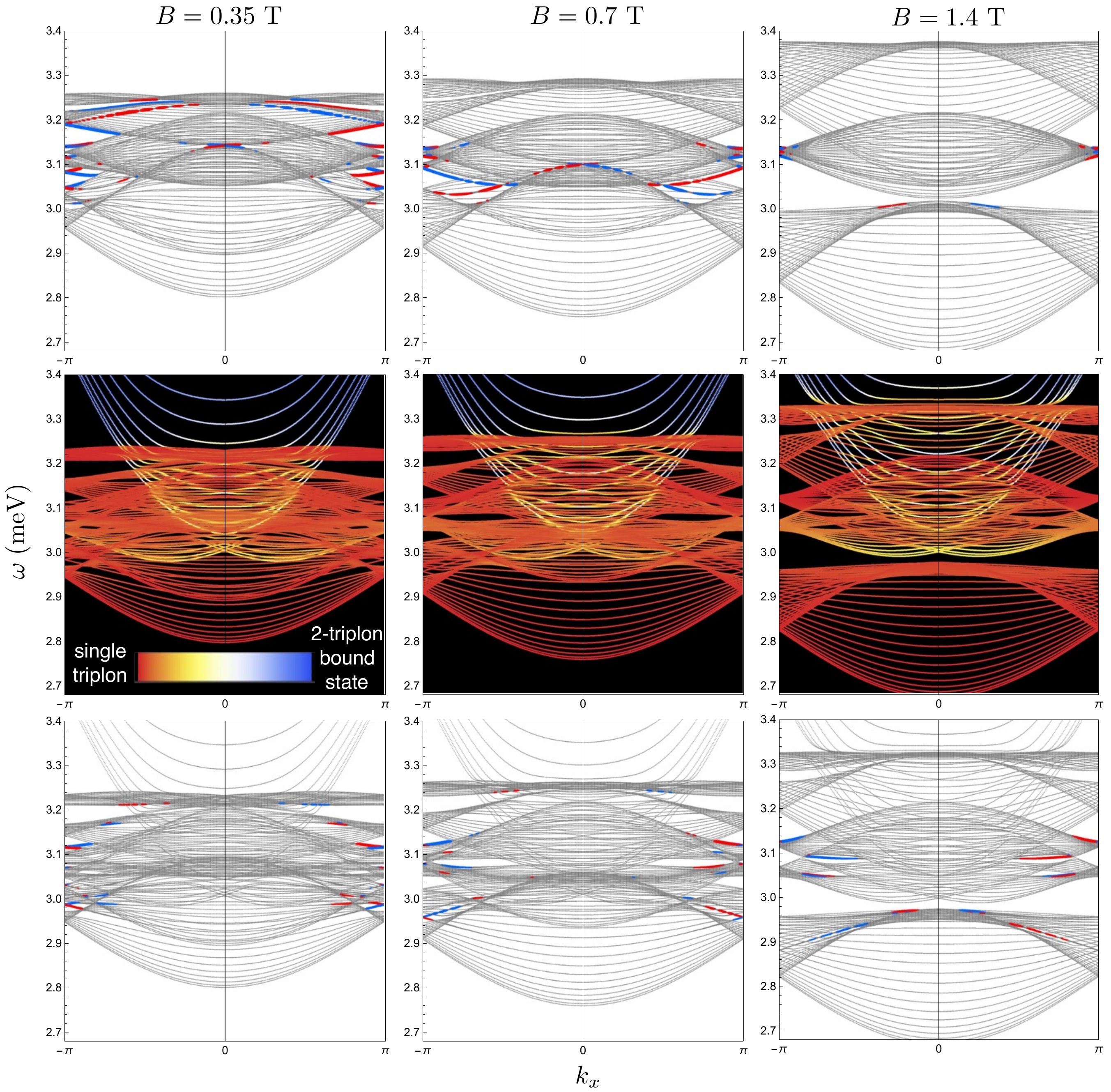}
\caption{ 
{\bf Excitation spectra and edge states of a sample with a strip geometry for different values of the magnetic field.} The top row shows the spectra for our model without the bound state. Modes that are localized near the top and bottom of the strip are colored in blue and red, respectively. The middle row shows the hybridization between the triplon excitations and the singlet bound state. The edge states are highlighted in the bottom row. 
\label{fig:edge_field}}
\end{figure}

From the Chern number calculation we have seen that because of the anisotropy and  the hybridization with the bound state, the magnetic excitation spectra contain various bands with 
non-trivial topology. As a function of field there exist various topological transitions involving different pairs of bands. This is also reflected by the rich structure of the thermal Hall effect. 
In Fig.~\ref{fig:edge_field} we present the complex field dependence of the edge states that are present on a strip of 20 rows of dimers. 

To identify edge states,  for each eigenvalue $\epsilon_i (k_x)$ we compute the overlap of the corresponding eigenstate with all single triplon and 2-triplon singlet states on a given layer $j$. From the resulting probability distribution $P_{i,k_x}(j)$ we can easily identify edge states from the condition that the mean $\langle j\rangle_{i,k_x}$ of the distribution lies within the top (blue) or bottom (red) two layers of the strip.

\section{Further Comparison between Theory and Experiment}

Our fitting procedure is as follows. We take various momentum cuts through the data and fit the peaks to gaussians with variable mean and variance taking the minimal number of gaussians necessary to obtain a good fit to each cut. In this way we obtain a set of points tracking the dispersion curves of the single triplon modes. To these points, we fit the single triplon model using a least square minimization algorithm and obtain a set of exchange parameters. The parameters are weakly correlated. For example, $J$ sets the $~3$ meV gap, the DM couplings have quite different effects on the dispersion, the further neighbor hopping term breaks the reflection symmetry about the central triplon mode and the $J_{\rm zz}$ and $J_{\rm xy}$ anisotropies break the zero field degeneracy at the $M$ point in different ways. Constant $\vert \boldsymbol{Q} \vert$ cuts indicate some degeneracy breaking at the $M$ point at zero field and hence that $J_{\rm zz}$ or $J_{\rm xy}$ or both are non-negligible. As a next step we use the full model including the bound state and allow the single triplon parameters to relax. 

\begin{figure}[htpb]
\includegraphics[width=0.7\columnwidth]{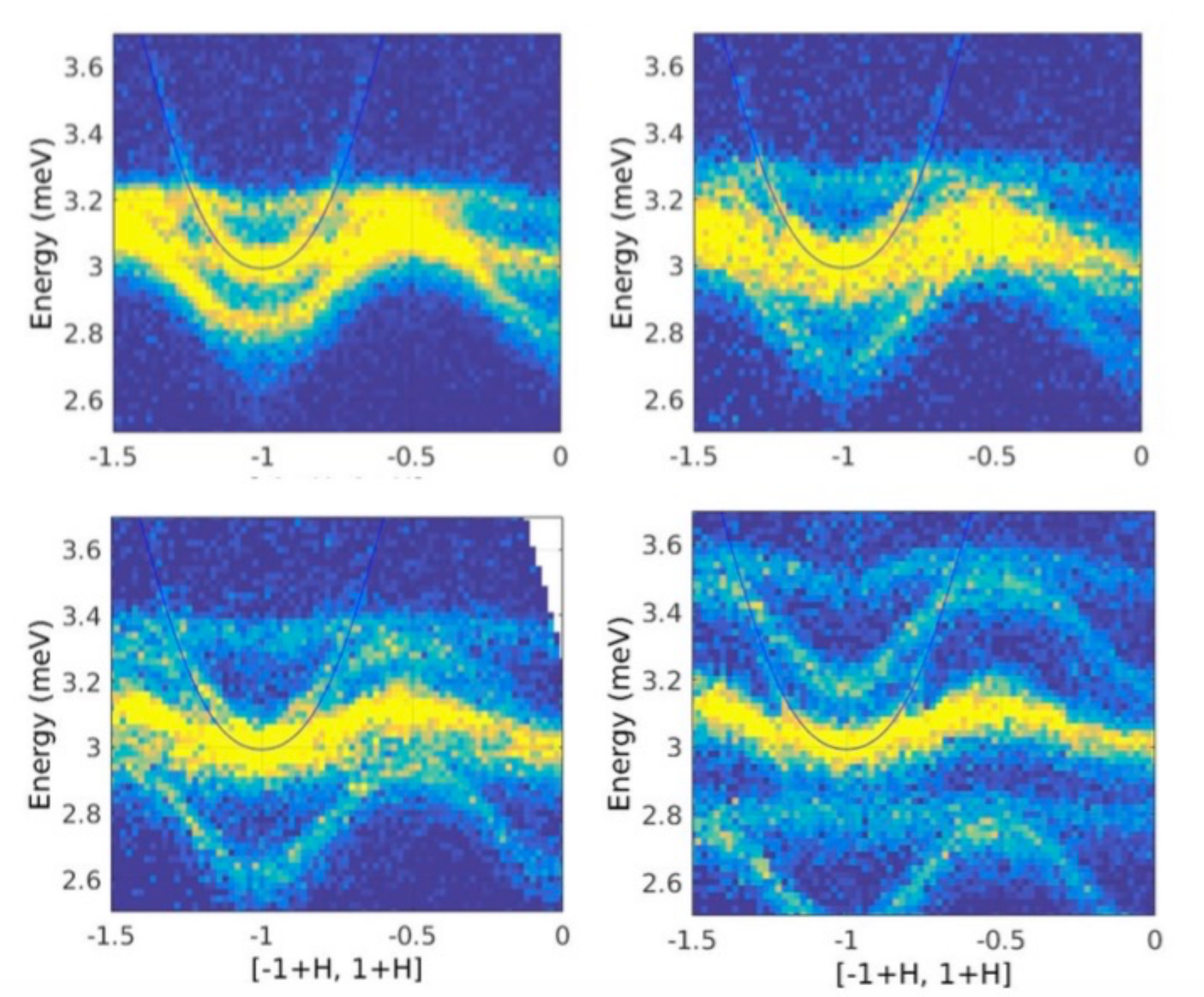}
\caption{ 
{\bf Field independence of singlet mode} Panels showing cuts along $[-1+H,1+H]$ in four different fields - in reading order $0$T, $0.7$T, $1.4$T and $2.8$T. A polynomial fit to the dispersive mode intersecting the singlet triplon excitations is identically plotted on top of these panels to illustrate the field independence of this mode.
\label{fig:S4}}
\end{figure}

\begin{figure}[htpb]
\centering
\hspace{3cm}
\centering
\includegraphics[width=\columnwidth]{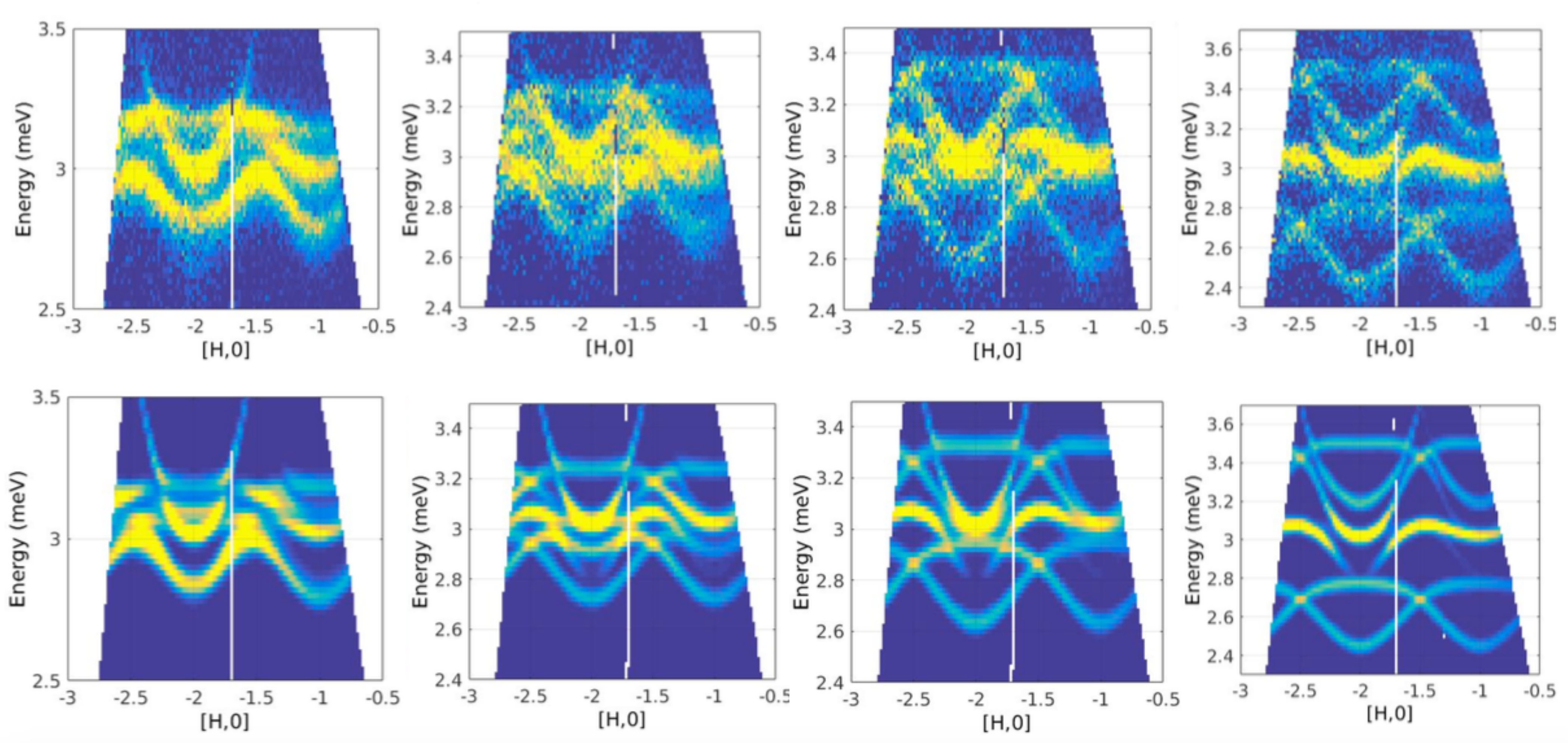}
\caption{ 
{\bf Experimental cuts along $[H,0]$ and corresponding dynamical structure factor of the model described in the main text.}
\label{fig:S5}}
\end{figure}

\begin{figure}[htpb]
\centering
\hspace{3cm}
\centering
\includegraphics[width=\columnwidth]{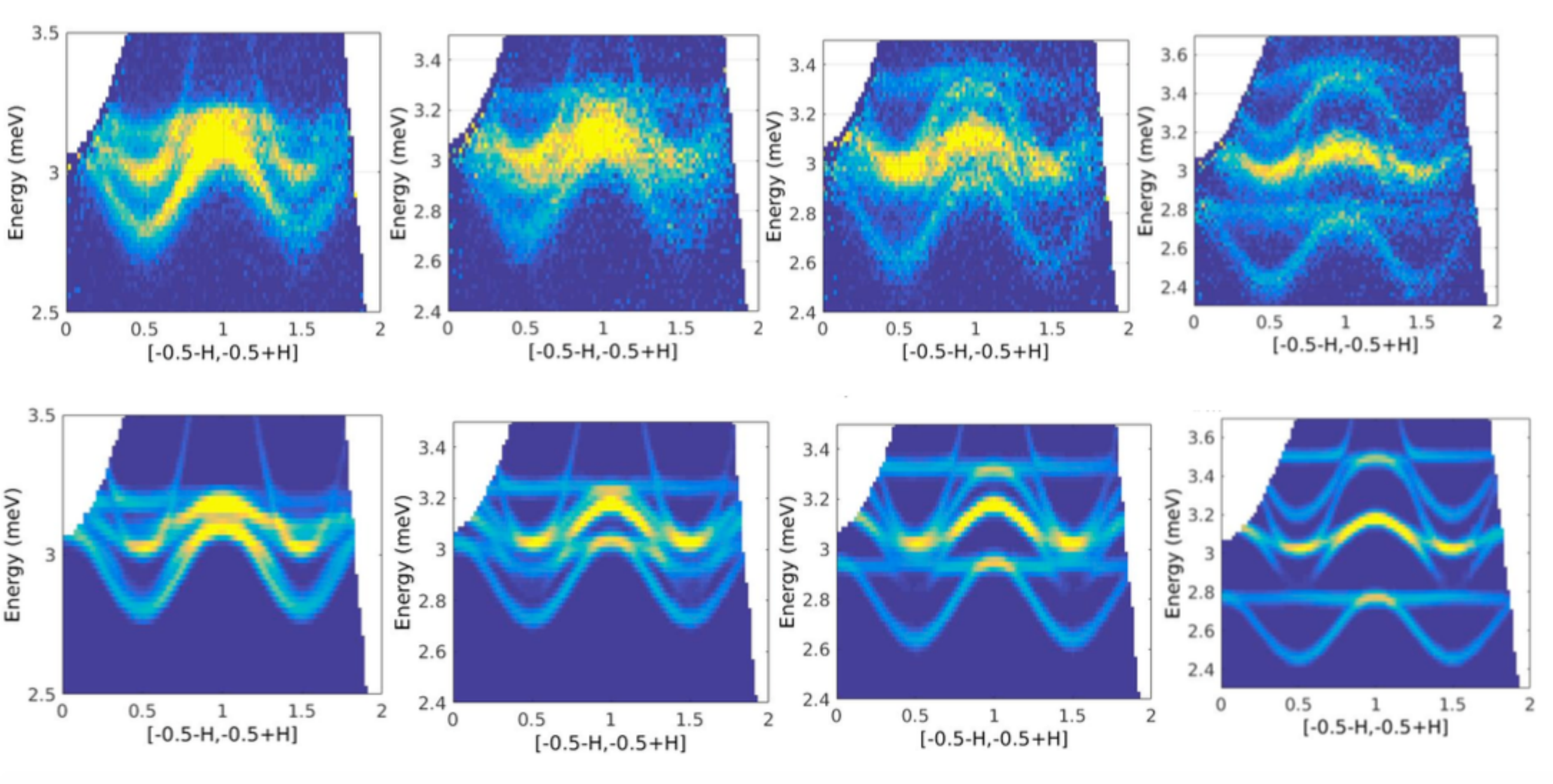}
\caption{ 
{\bf Experimental cuts along $[-0.5-H,-0.5+H]$ and corresponding dynamical structure factor of the model described in the main text.}
\label{fig:S6}}
\end{figure}

\begin{figure}[htpb]
\centering
\hspace{3cm}
\centering
\includegraphics[width=\columnwidth]{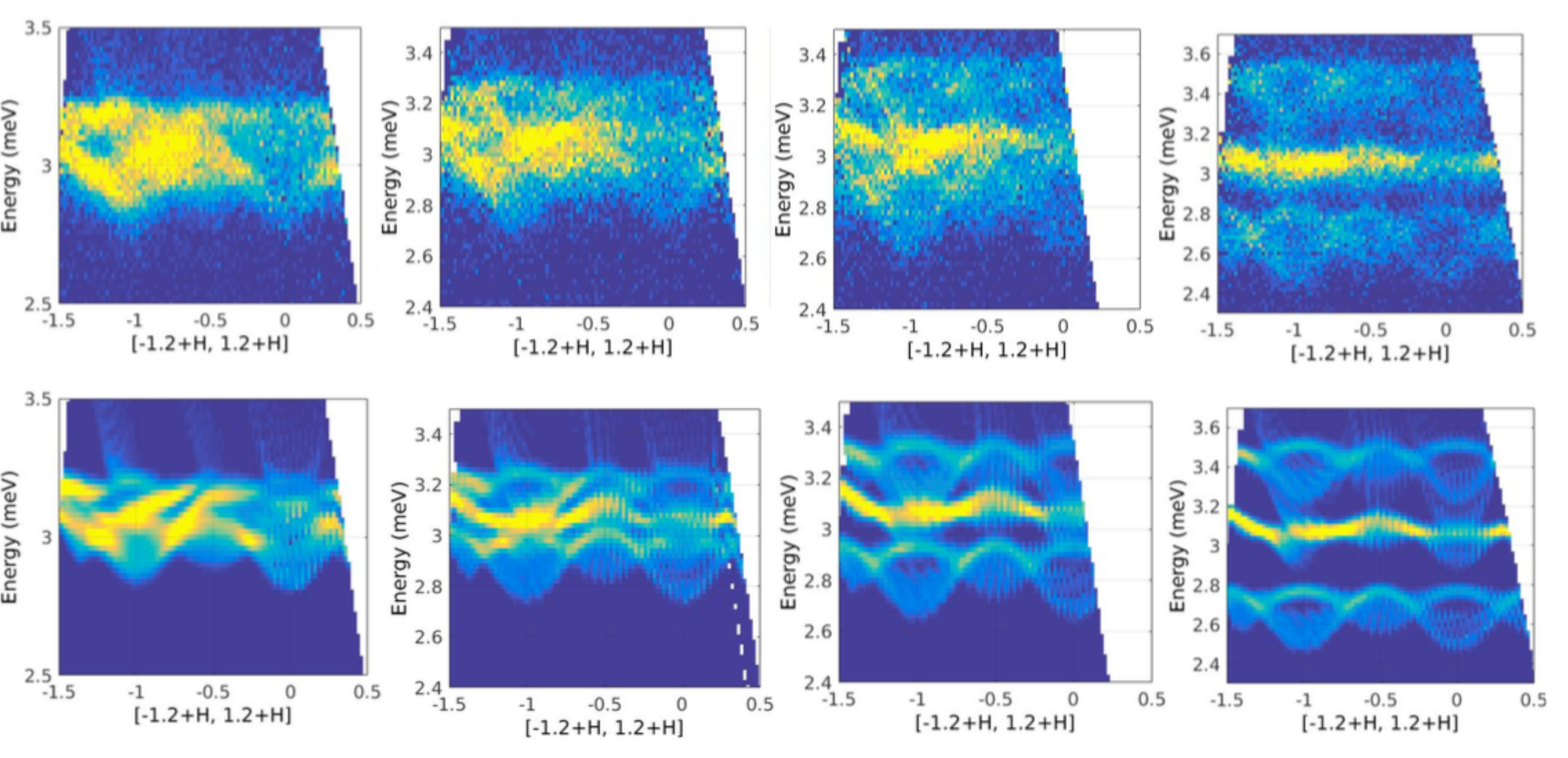}
\caption{ 
{\bf Experimental cuts along $[-1.2+H,-1.2+H]$ and corresponding dynamical structure factor of the model described in the main text.}
\label{fig:S7}}
\end{figure}

Here we present a sequence of figures showing the experimental inelastic neutron scattering intensity at $0$, $0.7$ T, $1.4$ T and $2.8$ T for different cuts through momentum space. These are compared with the best fits obtained from our full model. The comparison illustrates that both dispersions and intensities are captured by the model. In addition to the $[-1+H,1+H]$ cut in the main text, we show here $[H,0]$ (Fig.~\ref{fig:S4}), $[-0.5-H,-0.5+H]$ (Fig.~\ref{fig:S5}), $[-1.2+H,-1.2+H]$ (Fig.~\ref{fig:S6}) and $[-1.5,H]$ (Fig.~\ref{fig:S7}). The field independence of mode X, which we have identified as an $S=0$ bound state of two triplons, is most clearly visible in Fig.~\ref{fig:S4}.

\begin{figure}[htpb]
\centering
\hspace{3cm}
\centering
\includegraphics[width=\columnwidth]{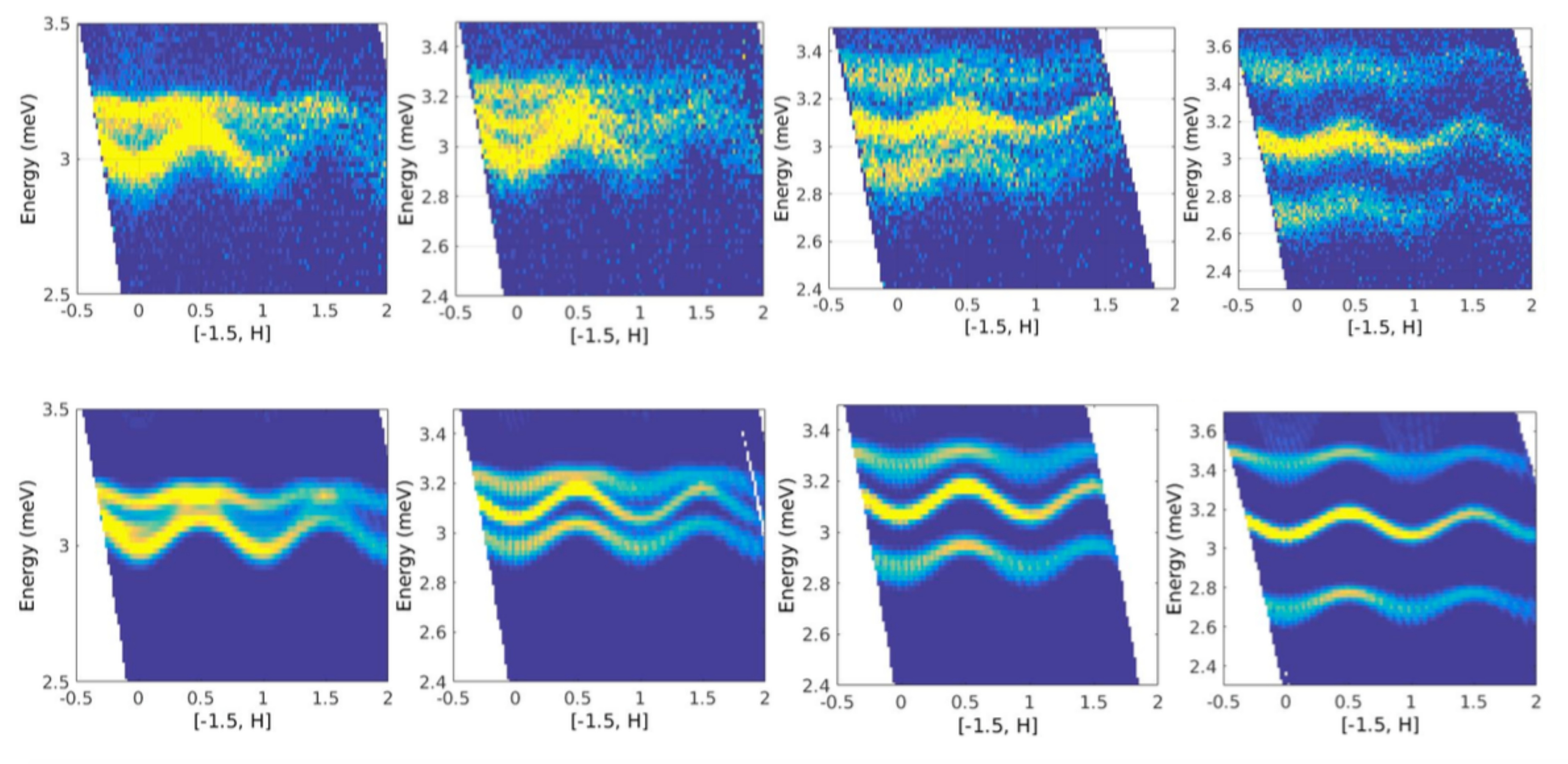}
\caption{ 
{\bf Experimental cuts along $[-1.5,H]$ and corresponding dynamical structure factor of the model described in the main text.}
\label{fig:S8}}
\end{figure}


\begin{thebibliography}{99}
\bibitem{remark} Amusingly, the DM exchange gives a vertex for the decay of an $S=0$ state into an $S=1$ state. The coupling is not rotationally symmetric and hence does not conserve angular momentum on its own. The angular momentum is taken up by the lattice highlighting the importance of spin-orbit coupling to the existence of antisymmetric exchange.
\end{thebibliography}
\end{document}